\documentclass[acmsmall,screen,table]{acmart}

\AtBeginDocument{%
  }

\setcopyright{acmlicensed}
\copyrightyear{2025}
\acmYear{2025}
\acmDOI{XXXXXXX.XXXXXXX}

\acmISBN{978-1-4503-XXXX-X/18/06}

\usepackage{cleveref}
\usepackage[linesnumbered,ruled,vlined]{algorithm2e}
\usepackage{multirow}

\newcommand{\toolname}[0]{\textit{PLocator}}
\newcommand{\tgt}[0]{$f_{tgt}$}

\newcommand{\cp}[1]{\textcolor{black}{#1}}

\renewcommand{\paragraph}[1]{\vskip 0.05in \noindent {\it #1.}}

\usepackage{tikz}
\usepackage{pifont}
\usepackage{subcaption}
\usepackage{enumitem}
\usepackage{tcolorbox}
\usepackage{multirow}
\usepackage{threeparttable}
\usepackage{adjustbox}
\usepackage{enumitem}
\usepackage{hyperref}

\crefname{section}{\S}{\S\S}
\Crefname{section}{\S}{\S\S}

\definecolor{lightblue}{RGB}{0, 127, 255}
\definecolor{lightorange}{RGB}{255, 153, 51}
\definecolor{lightpurple}{RGB}{255, 102, 255}
\definecolor{lightgreen}{RGB}{204, 255, 204}

\definecolor{darkred}{RGB}{204, 0, 0}

\begin{document}

\title{Fine-Grained 1-Day Vulnerability Detection in Binaries via Patch Code Localization}

\author{Chaopeng Dong}
\authornote{Work is done at Nanyang Technological University.}
\email{dongchaopeng@iie.ac.cn}
\orcid{0009-0000-4729-7778}
\affiliation{%
  \institution{Beijing Key Laboratory of IOT Information Security Technology, Institute of Information Engineering, CAS; School of Cyber Security, University of Chinese Academy of Sciences}
  \city{Beijing}
  \state{Beijing}
  \country{China}
}

\author{Jindong Guo}
\affiliation{%
  \institution{Institute of Information Engineering, CAS; School of Cyber Security, University of Chinese Academy of Sciences}
  \city{Beijing}
  \country{China}}
\email{guojingdong@iie.ac.cn}

\author{Shouguo Yang}
\affiliation{%
  \institution{Zhongguancun Laboratory}
  \city{Beijing}
  \country{China}}
\email{yangshouguo@outlook.com}

\author{Yang Xiao}
\affiliation{%
  \institution{Institute of Information Engineering, CAS; School of Cyber Security, University of Chinese Academy of Sciences}
  \city{Beijing}
  \country{China}}
\email{xiaoyang@iie.ac.cn}

\author{Yi Li}
\affiliation{%
  \institution{Nanyang Technological University}
  \city{Singapore}
  \country{Singapore}}
\email{yi_li@ntu.edu.sg}

\author{Hong Li}
\affiliation{%
  \institution{Institute of Information Engineering, CAS; School of Cyber Security, University of Chinese Academy of Sciences}
  \city{Beijing}
  \country{China}}
\email{lihong@iie.ac.cn}

\author{Zhi Li}
\affiliation{%
  \institution{Beijing Key Laboratory of IOT Information Security Technology, Institute of Information Engineering, CAS; School of Cyber Security, University of Chinese Academy of Sciences}
  \city{Beijing}
  \country{China}}
\email{lizhi@iie.ac.cn}

\author{Limin Sun}
\authornote{Corresponding Author}
\affiliation{%
  \institution{Beijing Key Laboratory of IOT Information Security Technology, Institute of Information Engineering, CAS; School of Cyber Security, University of Chinese Academy of Sciences}
  \city{Beijing}
  \country{China}}
\email{sunlimin@iie.ac.cn}
\renewcommand{\shortauthors}{Dong et al.}

\begin{abstract}
\cp{1-day vulnerabilities in binaries have become a major threat to software security. Patch presence test is one of the effective ways to detect the vulnerability. However, existing patch presence test works do not perform well in practical scenarios due to the interference from the various compilers and optimizations, patch-similar code blocks, and irrelevant functions in stripped binaries. In this paper, we propose a novel approach named \toolname{}, which leverages stable values from both the patch code and its context, extracted from the control flow graph, to accurately locate the real patch code in the target function, offering a practical solution for real-world vulnerability detection scenarios.}

\cp{
To evaluate the effectiveness of PLocator, we collected 73 CVEs and constructed two comprehensive datasets ($Dataset_{-irr}$  and $Dataset_{+irr}$), comprising 1,090 and 27,250 test cases at four compilation optimization levels and two compilers with three different experiments, i.e., Same, XO (cross-optimizations), and XC (cross-compilers). The results demonstrate that \toolname{} achieves an average TPR of 88.2\% and FPR of 12.9\% in a short amount of time, outperforming state-of-the-art approaches by 26.7\% and 63.5\%, respectively, indicating that \toolname{} is more practical for the 1-day vulnerability detection task.
}
\end{abstract}



\ccsdesc[500]{Security and privacy~Software security engineering}
\keywords{Vulnerability Detection, Patch Presence Test, Binary Code Similarity Detection}


\maketitle

\section{Introduction}\label{sec: introduction}
1-day (known) vulnerabilities have emerged as a significant threat in software systems due to the
widespread adoption of open-source software (OSS).
According to the CVE data~\cite{cvedetail}, the number of confirmed vulnerabilities has surged from 7,928 to 29,065 over the past decade.
As highlighted in Synopsys's report~\cite{ossRiskSynopsys}, 89\% of the codebases incorporate open-source code that has been stagnant for at least four years and components that have not received updates for two years, rendering them highly susceptible to 1-day vulnerabilities.

Given a known vulnerability, e.g., a vulnerable function, the task of \emph{1-Day Vulnerability Detection} aims to locate all homologous vulnerable functions within the target binary (which can be viewed as a set of functions).
The difficulty of identifying 1-day vulnerabilities lies in the potentially enormous amount of irrelevant functions within the target binary.
To accomplish this task, an efficient and effective technique called \emph{Binary Code Similarity Detection} (BCSD)~\cite{Sbjrnsen2009DetectingCC} has been proposed, which compares two or more pieces of binary code to identify their similarities and differences~\cite{haq2019survey}.
Given a reference vulnerable function $f_{vul}$, BCSD retrieves potentially vulnerable functions from a large codebase efficiently with high recall. 
However, it struggles to distinguish vulnerable functions from fixed ones since the differences
between them are often subtle~\cite{Zhao2020PatchScopeMO}, which leads to false positives.   
Thus, \emph{Patch Presence Test}~\cite{Zhang2018PreciseAA} is proposed as a follow-up procedure, to address the limitation of BCSD. Its input normally includes a patch file, two reference functions (i.e., vulnerable $f_{vul}$ and fixed $f_{fix}$) of the project, and a set of functions to be tested. The features generated from the patch determine whether a known security patch has been applied to the target binary.

\begin{figure*}[ht]
  \centering
  \includegraphics[width=\linewidth]{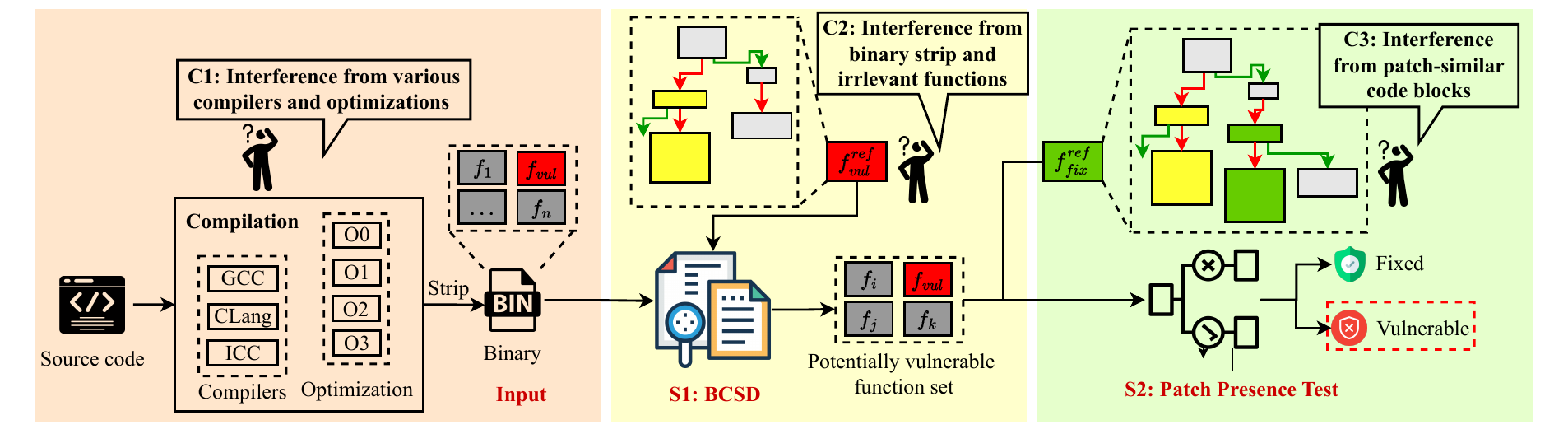}
  \caption{Workflow of the 1-day vulnerability detection task in stripped binary. The functions in the red, green grey rectangles refer to the vulnerable, fixed, and irrelevant functions, respectively. The yellow code blocks are similar to the green patch blocks.}
  \label{fig: vul_search_task}
\end{figure*}

\paragraph{1-Day Vulnerability Detection Workflow}
\Cref{fig: vul_search_task} illustrates the 1-day vulnerability detection task workflow based on BCSD and \emph{patch presence test}.
The input is a stripped binary (i.e., a set of functions) compiled from the source code with various compilers and optimizations. 
Initially, BCSD identifies the top-k most similar functions in the binary as potentially vulnerable by matching them to the reference vulnerable function $f_{vul}^{ref}$. Subsequently, \emph{patch presence test} classifies each of the retrieved functions into two categories: vulnerable or fixed, by verifying the presence of the patch code. 


\paragraph{Existing Approaches and Challenges}
The current techniques for \emph{patch presence test} still suffer from many limitations when integrated into the vulnerability detection workflow.
We roughly divide the existing \emph{patch presence test} techniques into \emph{syntactic-based} and \emph{semantic-based} approaches.
Syntactic-based approaches~\cite{Zhang2018PreciseAA,Xu2020PatchBV,Xu2023PatchDiscoveryPP} concentrate on the syntactic modifications made in patches, such as the topology of the CFG and normalized instructions in basic blocks. These approaches are efficient, owing to the lightweight comparison over the syntactic features.
Semantic-based approaches~\cite{Jiang2020PDiffSP, Yang2023TowardsPB, Zhan2023PS3PP} attempt to capture the semantic information by simulating the program's execution with existing techniques, such as symbolic execution. During the execution, the side effects (e.g., memory status and function call) are recorded as semantic information of the target, which is then compared with the references to determine the patch presence.
However, the emulation process is time-consuming, making it unsuitable for large-scale testing.
Despite the recent advances, three challenges remain in \emph{patch presence test} in binaries. 

\begin{itemize}[leftmargin=.25in]
    \item[\textbf{C1}] \cp{\textbf{Interference from various compilers and optimizations.}} 
    Developers frequently utilize diverse compilers and optimizations to satisfy specific requirements when producing binaries, resulting in numerous false negatives for syntactic-based methods. This occurs because the binary code generated with different compilers and optimizations diverges significantly~\cite{Chen2013TheIO}. Syntactic-based methods cannot distinguish whether the differences in binary code are due to compilation settings or patches.
    \item[\textbf{C2}] \cp{\textbf{Interference from binary strip and irrelevant functions.}
    Most of the \emph{patch presence test} approaches assume that the binary contains function symbols~\cite{Zhan2023PS3PP, Jiang2020PDiffSP, Zhang2018PreciseAA} or the BCSD tool can retrieve the target functions precisely~\cite{Xu2020PatchBV} so that they only consider the difference between vulnerable and fixed functions and treat the \emph{patch presence test} as a binary classification task (i.e., vulnerable or fixed). However, we argue that such a hypothesis is problematic, given the fact that for real-world binaries they often strip the symbols and the BCSD tool reports many false positives when the size of the function pool is large~\cite{Wang2022jTransJT, Yang2023AsteriaProED}, as discussed in a preliminary study on BCSD~\Cref{sec: preliminary study on BCSD}.  
    As such, when the input functions are irrelevant to the vulnerability, these \emph{patch presence test} approaches still classify them as vulnerable or fixed (grey rectangles in the ``Potentially vulnerable function set'' section of \Cref{fig: vul_search_task}), which have little meaning and cause many false positives. 
    }
    \item[\textbf{C3}] \cp{\textbf{Interference from patch-similar code blocks.}
    Code blocks that are similar or identical are common in binaries. For example, the same error handling code (e.g., throwing exceptions or printing errors) may be reused to fix similar vulnerabilities~\cite{APIDesignCPP}. Therefore, focusing on the patch code without considering the context and the relationship between them may lead to misidentification (i.e., taking vulnerable functions with patch-similar code as fixed), even for semantic-based methods.}
    
\end{itemize}

\paragraph{Our Approach}
To address the challenges, we propose \toolname{} (\underline{P}atch \underline{Locator}), an innovative approach that can accurately and efficiently detect 1-day vulnerabilities in binaries.

To tackle \textbf{C1}, we use the stable values extracted from the two types of key instructions (condition comparison and function call) as anchors, which are stable across compilers and optimizations. 
To distinguish the key instructions with the same values, we also extract the constant values associated with the key instruction through data flow dependencies as auxiliary information. We generate the signatures based on the extracted anchors to reduce the interference from various compilers and optimizations.

To tackle \textbf{C2}, we first filter irrelevant functions at a coarse-grained level by comparing the matched unique anchor values between the reference functions and target functions. Subsequently, we classify the target functions as irrelevant by examining whether the patch code and its context are matched or not.

To tackle \textbf{C3}, we utilize the context of the patch code to 
eliminate patch-similar code blocks (e.g., yellow blocks mentioned in~\Cref{fig: vul_search_task}) by verifying whether the matched patch code and context code satisfy the original control flow relationship in patch.

\paragraph{Evaluation}
\cp{To evaluate the effectiveness of our work from multiple perspectives, we first collected 73 CVEs and 112 distinct vulnerable functions from four popular projects. And then, we constructed two datasets: $Dataset_{-irr}$ (without irrelevant functions) and $Dataset_{+irr}$ (with irrelevant functions), based on the collected data and compiled binaries at four different optimization levels (i.e., O0 to O3) and two compilers (i.e., GCC and Clang). The total number of test cases in the two datasets are 1,090 and 27,250, respectively. We conducted three experiments, i.e., Same, XO (cross-optimizations), and XC (cross-compilers), on the two datasets to evaluate the ability of three state-of-the-art approaches and \toolname{} to identify vulnerable, fixed, and irrelevant functions.
The experimental results demonstrate that \toolname{} outperforms state-of-the-art approaches on TPR and FPR by a large margin. \toolname{} achieves an average TPR and FPR of 87.9\% and 16.1\% on $Dataset_{-irr}$, 88.5\% and 9.6\% on $Dataset_{+irr}$, outperforming the second-best approaches by 24.8\% (TPR) and 52.1\% (FPR) on $Dataset_{-irr}$, 28.4\% (TPR) and 75.0\% (FPR) on $Dataset_{+irr}$, respectively. \toolname{} completes the detection process in a short amount of time, averaging 0.14 seconds per test case.
}

\paragraph{Contributions}
\cp{
Overall, we summarize our contributions as follows: 
\begin{enumerate}[leftmargin=.25in]
    \item We systematically outline the workflow for the 1-day vulnerability detection task and present the limitations of existing patch presence test approaches through a preliminary study of the BCSD tool and a motivating example. 
    \item To address the challenges, we propose \toolname{}, a novel approach that leverages stable values in both patch code and its context to accurately locate the patch code and detect vulnerabilities, which is more practical for the 1-day vulnerability detection task. 
    \item We collect the original data from NVD, comprising 73 CVEs and 112 distinct vulnerable functions, and construct two datasets using four optimization levels and two compilers, generating a substantial number of test cases to evaluate the effectiveness of various patch presence test approaches across three experiments. The results demonstrate the effectiveness of \toolname{} in accurately and efficiently detecting 1-day vulnerabilities.
\end{enumerate}
}

\section{Background}
In this section, we begin with a preliminary study to reveal the limitations of BCSD. Next, we define key terms used throughout the paper for clarity and easy reference. Finally, we present our motivating example. 

\begin{figure}[ht]
  \centering
  \includegraphics[width=\linewidth]{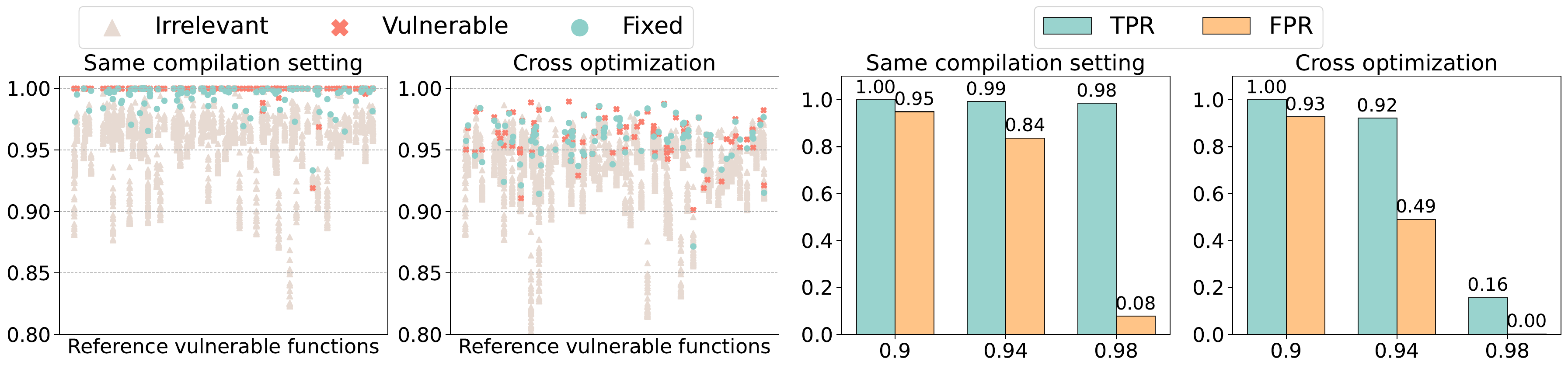}
  \caption{The binary function similarity distribution conducted by jTrans. The left two scatter plots show the BCSD similarity distribution from the binaries under the same compilation setting and across different optimizations, respectively. The x-axis represents different reference vulnerable functions, while the y-axis represents the similarities of top-50 most similar functions from the function pool with 6,000 candidates. The right two figures show the TPR and FPR on two compilation settings under different thresholds. The target function exceeds the threshold will be identified as ``vulnerable'', otherwise it is ``irrelevant''.}
  \label{fig: bcsd score dist}
\end{figure}

\subsection{Preliminary Study on BCSD}\label{sec: preliminary study on BCSD}
\cp{To expose the limitations of BCSD on 1-day vulnerability detection and the interference from the irrelevant functions for \emph{patch presence test}, we first conduct a preliminary study on jTrans, i.e., a state-of-the-art BCSD approach, by matching the reference vulnerable functions against a large function pool under the same compilation settings and across optimizations, respectively. The details of the experimental setup are presented in~\Cref{sec: experimental setup}. 
Then, we treat vulnerable functions as positive, fixed and irrelevant functions as negative, and employ TPR (True Positive Rate) and FPR (False Positive Rate) as the evaluation metrics, defined as follows:
\begin{equation}
    TPR = \frac{TP}{TP + FP},
    FPR = \frac{FP}{TN + FP},
\end{equation}
where $TP$, $FP$, and $FN$ refer to the number of vulnerable functions truly classified as vulnerable, irrelevant and fixed functions erroneously classified as vulnerable, and vulnerable functions erroneously classified as fixed or irrelevant, respectively. 
}

\cp{~\Cref{fig: bcsd score dist} presents the BCSD similarity distribution and metrics under different BCSD thresholds. From the similarity distribution, it is evident that jTrans successfully recalled the vulnerable functions but mistakenly included many fixed and irrelevant functions as the potentially vulnerable function set in~\Cref{fig: vul_search_task}. A higher threshold reduces false positives but also lowers the TPR a lot, which is unacceptable. Therefore, it is crucial for the \emph{patch presence test} approach to accurately distinguish fixed and irrelevant functions from vulnerable ones as a follow-up work of BCSD.
}

\subsection{Terms and Definitions}
\cp{\noindent\textbf{Patch and reference functions.} A \emph{patch} fixes a vulnerability, which can be represented as a source ``diff'' file. The patch code includes vulnerable (deleted) code and fixed (added) code, which are separated in two reference functions: vulnerable function $f_{vul}^{ref}$ and fixed function $f_{fix}^{ref}$. These reference functions are employed to extract features for the patch presence test.
}

\noindent\textbf{Anchor ($ac$).} An anchor is a stable data structure that persists across compilers and optimizations in our study, comprising three elements: value, type, and auxiliary information. 

\noindent\textbf{Auxiliary information ($aux$).} To differentiate anchors with the same value and type, we additionally extract numeric and string constants associated with the anchor through data dependencies as auxiliary information, denoted as $aux = \langle (c_1, t_1), \ldots, (c_n, t_n) \rangle$, where $c_i$ represents the constant and $t_i$ denotes its type.

\noindent\textbf{Anchor graph ($AG$).} This refers to a directed graph consisting of anchors that follow the control flow of the CFG (Control flow graph), denoted as $AG = (V, E)$, where $V = \{ac_1, \ldots, ac_n\}$ represents the set of anchors, and $E = \{(ac_i, ac_j) \mid ac_i \in V, ac_j \in V\}$ denotes the set of edges connecting the anchors.

\begin{figure*}[ht]
  \centering
  \includegraphics[width=\linewidth]{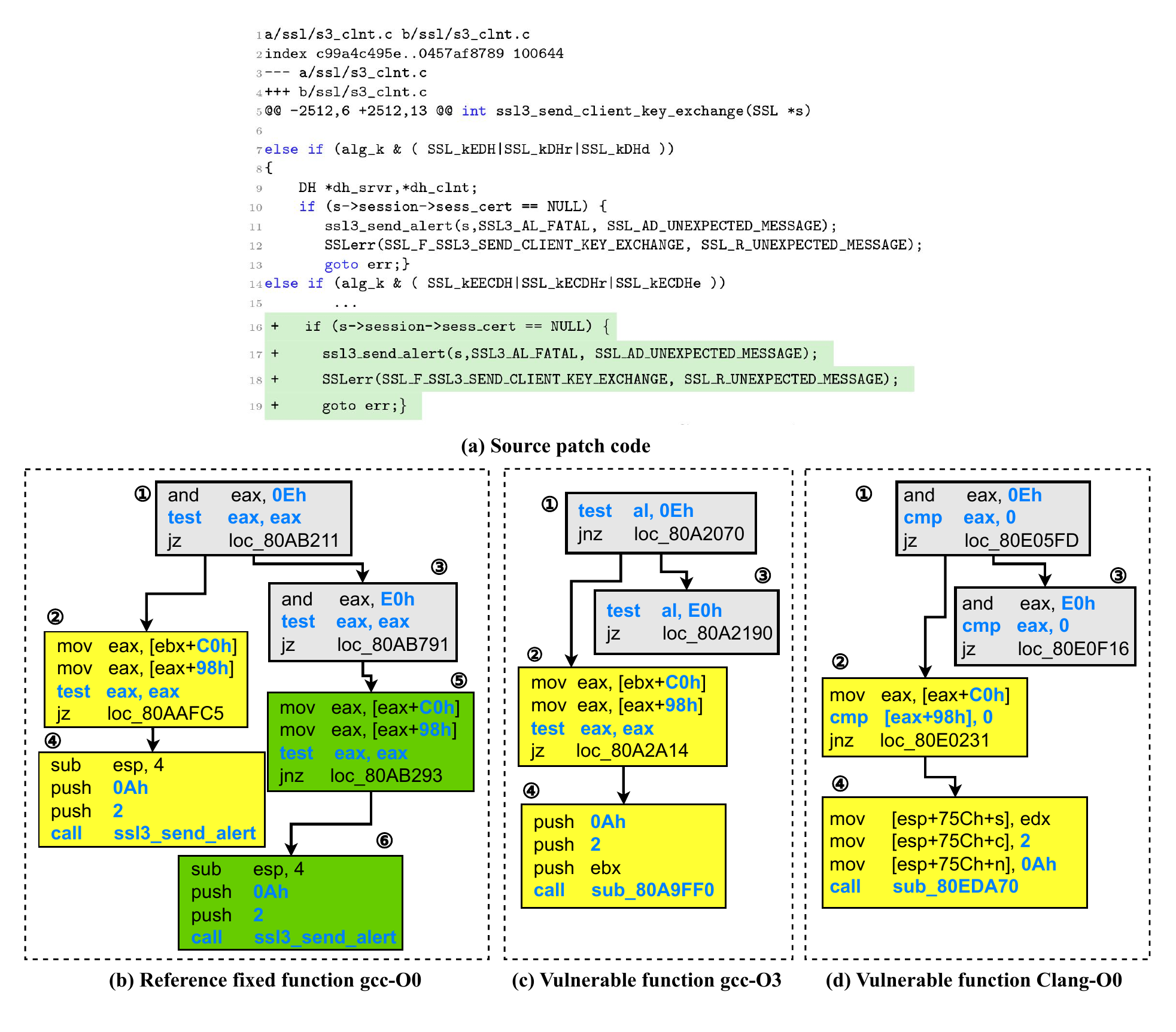}
  \caption{An motivating example of CVE-2014-3470. The yellow code blocks have similar assembly code as the green patch code blocks. The stable values and instructions are highlighted with blue text.}
  \label{fig: motivation}
\end{figure*}

\subsection{Motivating Example}\label{sec: motivating example}
We use CVE-2014-3470~\cite{cve-2014-3470}, a vulnerability of the famous cryptographic protocols software OpenSSL, to motivate our approach, as depicted in \Cref{fig: motivation}.~\Cref{fig: motivation}a is the source patch code which adds a check on whether variable \texttt{``s->session->sess\_cert''} is $NULL$, along with function call to error processing function \texttt{``ssl3\_send\_alert''}. 
~\Cref{fig: motivation}b refers to the CFG of fixed reference function compiled with gcc-O0, while ~\Cref{fig: motivation}b and ~\Cref{fig: motivation}c refer to the two CFG variants of vulnerable function compiled with gcc-O3 and Clang-O0, respectively. The yellow blocks in vulnerable functions present similar code to the green patch blocks in the fixed function~\Cref{fig: motivation}b. 

\paragraph{Limitations of Existing Approaches}
\cp{When detecting patches in the two variants, we found that both types of patch presence test approaches exhibit limitations.
\begin{enumerate}[leftmargin=.25in]
    \item \emph{Syntactic-based approaches, such as BinXRay~\cite{Xu2020PatchBV}, struggle to discern whether changes are introduced by the patch or compiler options.} For example, for the same source code statement (line 14 in ~\Cref{fig: motivation}a), reference fixed function in~\Cref{fig: motivation}b presents an \emph{and} operation (\texttt{``and, eax, E0H''}) before the comparison of the register \texttt{eax} while vulnerable function in ~\Cref{fig: motivation}b and ~\Cref{fig: motivation}c compare the register with constant directly (\texttt{``test al, E0h''}) and use different comparison instruction (\texttt{``cmp, eax, 0''}), respectively. These changes are irrelevant to the patch code and will lead to the misidentification of syntactic-based methods.  
    \item \emph{Semantic-based approaches, such as PS3~\cite{Zhan2023PS3PP} and Pdiff~\cite{Jiang2020PDiffSP}, fail to distinguish patch-similar code from the actual patch code due to their identical semantic features, resulting in misidentification.}. For example, the two yellow code blocks in~\Cref{fig: motivation}b present similar assembly code as the green code blocks (i.e., patch). Methods that extract semantic features from the green code blocks mistakenly identify the yellow blocks as the patch code, leading to the incorrect classification of the two variants as fixed.
\end{enumerate}
}


\paragraph{Observations and Illustration}
We present two major observations to address the aforementioned limitations.
First, by comparing the CFGs of three functions carefully, we found that two types of code (condition comparison and function call) and the constants associated with them through data dependencies remain consistent across compilers and optimizations, highlighted in \textcolor{lightblue}{\textbf{bold blue text}} (\textbf{Observation 1}). 
Second, by inspecting the green (patch) and yellow blocks in~\Cref{fig: motivation}b, we found that their distinguishing factor lies in their prerequisites, wherein one involves an \texttt{and} operation with \textcolor{lightblue}{\textbf{0Eh}}, while the other is \textcolor{lightblue}{\textbf{E0h}} (\textbf{Observation 2}).
These two observations inspire us to leverage the patch and its context within stable values to locate the patch code and verify the patch presence. We illustrate the process of \toolname{} for vulnerability detection as follows:

\textbf{Signature generation.}
To determine whether the two variants contain patch code or not, we first locate the patch code in~\Cref{fig: motivation}b with debug information and extract the stable values from it (i.e., bold blue text in blocks \textcircled{5} and \textcircled{6}). Subsequently, we further extract the stable values in blocks \textcircled{1} and \textcircled{3} of the patch as the context.

\textbf{Vulnerability detection.} Initially, we match the stable values of the patch and its context in two variants (i.e., \textcolor{lightblue}{\textbf{bold blue text}}). For function calls without symbols (i.e., \textcolor{lightblue}{\textbf{sub\_80A9FF0}}), we match them through BCSD by limiting the search space to a small range to reduce FPR. More details are discussed in ~\Cref{sec: anchor path matching}. 
Afterward, the candidate patch code (i.e., yellow blocks) will be filtered out since it does not satisfy the original control flow relationship between the patch and its context, i.e., patch block \textcircled{2} does not follow the execution after context block \textcircled{3}). Thus, we determine the two variants are vulnerable.

\section{Methodology}

\subsection{Task definition}
To be more precise, we define the task of \emph{1-Day Vulnerability Detection} as follows:

\begin{definition}[1-Day Vulnerability Detection Task]\label{srvd task}
Given a patch and two reference functions, i.e., $f_{vul}^{ref}$ and $f_{fix}^{ref}$, of a known vulnerability.
Let $P = \{f_1, f_2, \ldots, f_{n}\}$ be a pool of binary functions, where the binary function pool may be compiled with different compilers and optimizations from the reference functions.
The task aims to identify the vulnerable function among numerous irrelevant functions in $P$ by classifying the functions into vulnerable, fixed, or irrelevant.
\end{definition}

\subsection{Overview}
The goal of \toolname{} is to locate the patch code in a set of binary functions (function pool) with the given patch and two reference functions, and then determine the categories of the target functions. 

\begin{figure*}[ht]
  \centering
  \includegraphics[width=\linewidth]{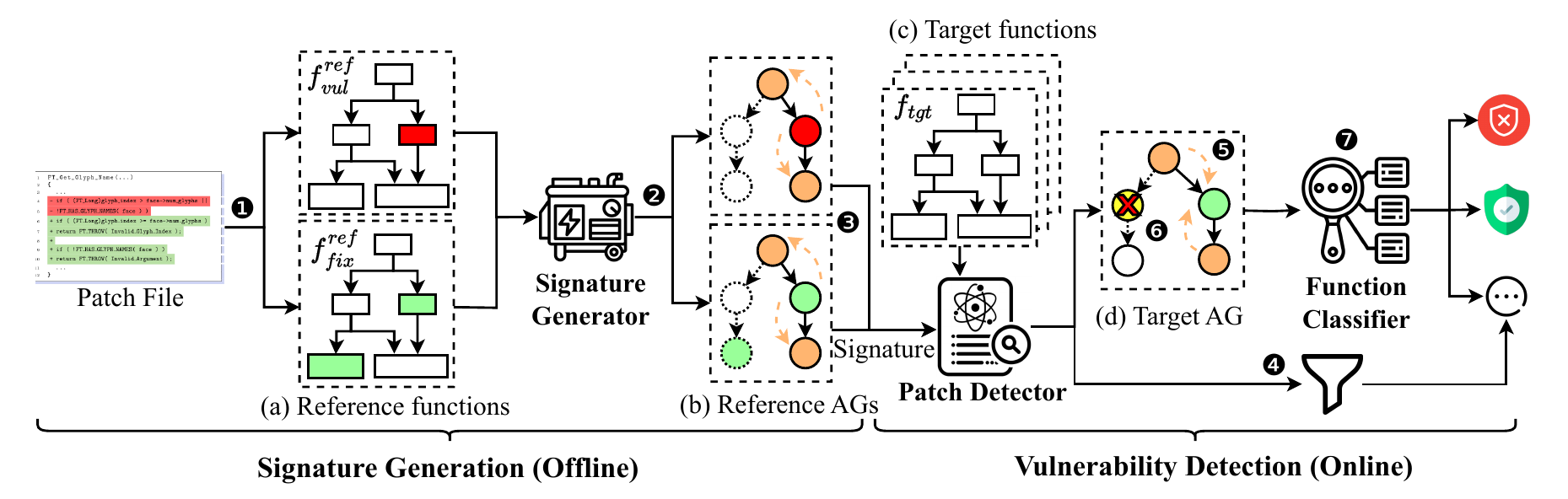}
  \caption{The workflow of \toolname{}, composed of 7 steps, which are patch code mapping(\ding{182}), anchor graph construction (\ding{183}), anchor path extraction
  (\ding{184}), irrelevant function filtering (\ding{185}), anchor path matching (\ding{186}), patch path verification (\ding{187}), and function classification (\ding{188}).}
  \label{fig: workflow}
\end{figure*}

\Cref{fig: workflow} depicts the workflow of \toolname{}, comprising two main stages: \emph{Signature Generation} (Offline) and \emph{Vulnerability Detection} (Online).
During the signature generation stage, \ding{182} the signature generator initially maps the source patch code into the two reference functions (i.e., red and green blocks in~\Cref{fig: workflow}a), then \ding{183} extracts the stable values as anchors from the CFGs to construct the anchor graph (AG) and \ding{184} select the most representative patch code (i.e., red and green nodes in~\Cref{fig: workflow}b) along with their context (i.e., orange nodes in~\Cref{fig: workflow}b) to form the signature. 

During the vulnerability detection stage, the anchor graphs of the target functions are constructed in the same way as in the previous stage. Subsequently, the patch detector \ding{185} filters the irrelevant functions based on the unique anchor values, and \ding{186} matches the patch code and its context in the target AG separately (i.e., yellow, green, and orange nodes in~\Cref{fig: workflow}d). Next, \ding{187} it eliminates the patch-similar code (i.e., the yellow node in the target AG~\Cref{fig: workflow}d) by verifying the control flow relationship between the patch and context code. Ultimately, \ding{188} the identified patch code and its context are sent to the function classifier to determine the categories of the target functions, which can be vulnerable, fixed, or irrelevant.

\subsection{Signature Generator}
The signature generator generates representative features of the patch code and its context code based on the patch file and two reference functions. The entire process comprises three steps: patch code mapping, anchor graph construction, and anchor path extraction.

\subsubsection{Patch Code Mapping} 
Patch code mapping aims to pinpoint the patch code in reference binary functions with the given patch file (i.e., deleted code in $f_{vul}^{ref}$ and added code in $f_{fix}^{ref}$) and debug information from the reference binaries. 

To achieve the object, we first standardize and compute the hash value~\cite{md5} of the source code lines in the patch file and two reference functions to minimize the influence of non-semantic modifications and improve the comparison efficiency, which is accomplished by eliminating all comments, braces, tabs, and white spaces, similar to previous studies~\cite{Xiao2020MVPDV, woo2021centris, Woo2022MOVERYAP}. Subsequently, we split the patch code into two parts (deleted and added), and map them to the source code of reference versions. Considering that identical code statements may recur within a function (e.g., \texttt{``goto error\;''}), we pinpoint the real patch code with its context source code in the patch file.   
Ultimately, for the mapped source code, we employ addr2line~\cite{addr2line}, a command-line tool that converts binary addresses into source file names and line numbers using debug information, to identify the patch blocks in two reference binary functions. 

\subsubsection{Anchor Graph Construction}\label{sec: anchor graph construction}
To create a signature that is stable across compilers and optimizations, we first construct the anchor graph from the CFG, which consists of anchors with auxiliary information attached. 
As illustrated in~\Cref{sec: motivating example}, we extract anchors from the two types of key instructions, \emph{condition comparison} and \emph{function call}, which heavily influence the control flow of the program and persist across compilers and optimizations.
The generation consists of two main steps as follows:

\begin{table}[]
\begin{adjustbox}{max width=\columnwidth}
\begin{threeparttable}

\caption{The types of key instructions and the corresponding anchor values.}
\label{tab: key_and_anchor}
\begin{tabular}{l|l|l|l|l|l}
\hline
\hline 
\textbf{Key instruction type} & \textbf{Key example} & \textbf{Anchor value and type} & \textbf{Sliced instruction type} & \textbf{Sliced example} & \textbf{Aux} \\ \hline
\multirow{2}{*}{Condition comparison} &
  \multirow{2}{*}{\texttt{cmp eax, 2}} &
  \multirow{2}{*}{\begin{tabular}[c]{@{}l@{}}Value: 2\\ Type: ``CMP''\end{tabular}} &
  Assignment &
  \texttt{mov eax, [eax + 0Eh]} &
  (0Eh, ``offset'') \\ \cline{4-6} 
                              &                      &                                & Arithmetic                       & \texttt{add eax, 2}     & (2, ``add'') \\ \hline
Function call &
  \texttt{call foobar} &
  \begin{tabular}[c]{@{}l@{}}Value: foobar\tnote{1}\\ Type: ``CALL''\end{tabular} &
  Parameter &
  \begin{tabular}[c]{@{}l@{}}\texttt{push 0Ah}\\ \texttt{mov [esp], 3}\end{tabular} &
  \begin{tabular}[c]{@{}l@{}}(0Ah, ``param'')\\ (3, ``param'')\end{tabular} \\ \hline 
\hline
\end{tabular}

\begin{tablenotes}
      \item[1] We do not utilize function names for matching when unavailable.
\end{tablenotes}
\end{threeparttable}
\end{adjustbox}

\end{table}

\begin{figure}[ht]
  \centering
  \includegraphics[width=0.9\linewidth]{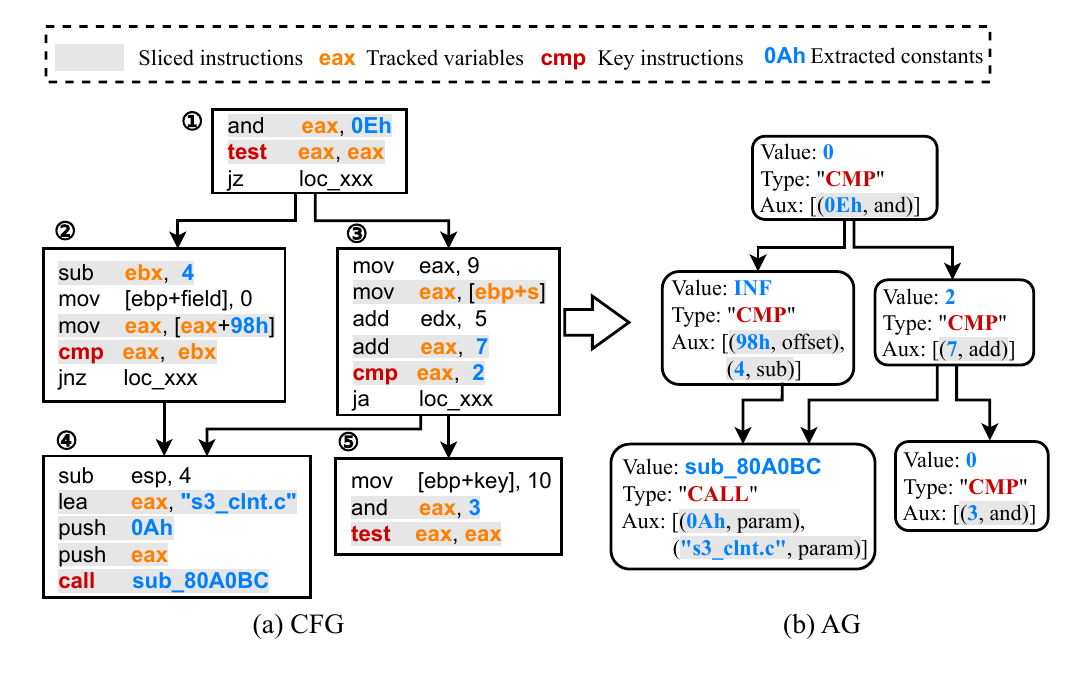}
  \caption{Example of anchor graph construction. The left and right graphs are the CFG and the generated AG, respectively. The dotted lines include the legends of the graph components. ``INF'' is a special flag for comparison between two different variables.}
  \label{fig: anchor graph construction}
\end{figure}

\cp{\textbf{\ding{182} Key instruction identification.} As shown in~\Cref{fig: anchor graph construction}a, we first identify two types of key instructions in CFG (e.g., \textcolor{darkred}{\textbf{cmp}}) based on the known instructions for specific operations (e.g., \texttt{test}, \texttt{cmp} for condition comparison and \texttt{call} for function call). After that, we extract the anchor value from the key instruction and determine the anchor type based on the key instruction type. The first three columns in~\Cref{tab: key_and_anchor} give two examples of the extracted anchor values and types. For other key instruction formats, such as \texttt{``test eax, eax''} for condition comparison (i.e., compare whether the variable is $0$ or ``NULL'' through \texttt{and} operation) and \texttt{``jmp loc\_80AAF63''} for indirect function call (i.e., where the jump target is the start address of a function), we convert them into equivalent key instruction formats as shown in column 2 of ~\Cref{tab: key_and_anchor} to extract the anchor values and types based on the predefined rules. 
\emph{Note that although we extract the anchor value of the function name for call instruction, we do not utilize it for matching directly when the function name is unavailable (e.g., ``sub\_80A0BC'')}.
}

\cp{\textbf{\ding{183} Backward instructions slicing.} Merely using constants in two types of key instructions is insufficient, as multiple anchors with identical values and types may exist in the same function. 
For example, two comparison instructions (\texttt{``test eax, eax''}) in block \textcircled{1} and \textcircled{5} in~\Cref{fig: anchor graph construction}a generate the same anchor value $0$.
The main difference between these two key instructions lies in the previous instructions associated through variable \textcolor{orange}{\textbf{eax}}, where one involves \emph{and} operation with \textcolor{lightblue}{\textbf{0Eh}}, while the other is \textcolor{lightblue}{\textbf{3}}. 
\emph{Intuitively, slicing techniques~\cite{Silva2012AVO} can be employed to incorporate relevant instructions and extract the associated constants}. Specifically, we can perform backward slicing on the blocks, track variables in the key instruction, and use them as the slicing criterion. 
For instance, we set the variables (\textcolor{orange}{\textbf{eax}} and \textcolor{orange}{\textbf{ebx}}) in \texttt{``cmp eax, ebx''} of block \textcircled{2} in~\Cref{fig: anchor graph construction}a as the slicing criterion and slice the instructions that are associated through data dependencies. The slicing results are highlighted with grey text background. From the sliced instructions, we update the tracking variables and extract the relevant constants as well as their types as the auxiliary information of the anchor. For function call instruction, we track the variables and extract constants from its parameters according to the calling conventions. In \emph{x86} architecture, the parameters are moved into the stack so that we slice the \texttt{push} instruction (e.g., \texttt{``push eax''}) and \texttt{mov} operation taking register \texttt{esp} as the target (e.g., \texttt{``mov [esp], 3''}). 
As shown in~\Cref{tab: key_and_anchor}, We roughly divide the sliced instruction types into three categories: assignment, arithmetic, and parameter. The last three columns show the sliced example and the corresponding extracted auxiliary information.
}

Ultimately, we extract the anchors along with their auxiliary information to construct the anchor graph (AG) by following the original control flow of the CFG, as illustrated in~\Cref{fig: anchor graph construction}b.

\begin{figure}[ht]
  \centering
  \includegraphics[width=0.8\linewidth]{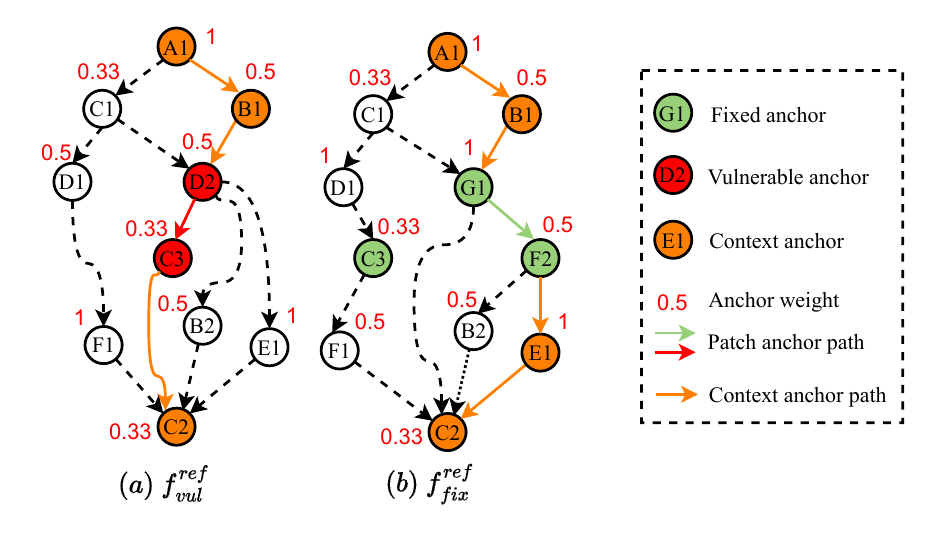}
  \caption{Example of anchor path extraction. The left and right graphs are the AG of $f_{vul}^{ref}$ and $f_{fix}^{ref}$, respectively.}
  \label{fig: anchor path extraction}
\end{figure}

\subsubsection{Anchor Path Extraction}\label{sec: anchor path extraction}
After constructing the anchor graph as the foundation, we need to further extract useful information that is related to the patch and its context for vulnerability detection. Therefore, we define the anchor path as the basic component to form the signature, which is a sequence of anchors following the control flow in $AG$, represented as $AP = \langle ac_1, \ldots, ac_m\rangle, ac_i \in AG$. We refer to the anchor path derived from the patch code as the patch (anchor) path, and the one sourced from the context code as the context (anchor) path, respectively. 

\cp{To explain the extraction process, we present an example as depicted in~\Cref{fig: anchor path extraction}. There are two AGs, i.e., AG for $f_{vul}^{ref}$ and AG for $f_{fix}^{ref}$ with anchor weights displayed in \textcolor{red}{red font}. 
The weight of each anchor in AG is calculated by $w_{ac} = \frac{1}{TF_{ac}}$, where $TF_{ac}$ denotes the term frequency of $ac$ (i.e., two anchors are considered identical when they share the same value, type, and auxiliary information).
Each anchor (node) is identified with a letter and a number (e.g., \textbf{C3}), and anchors with the same letter contain the same anchor values and auxiliary information. From $f_{vul}^{ref}$ to $f_{fix}^{ref}$, \textbf{F2} is newly introduced, \textbf{C3} is repositioned between \textbf{D1} and \textbf{F1}, and \textbf{D2} is transformed into \textbf{G1}, consequently altering the anchor weights. The whole extraction process consists of two main steps as follows:
}

\cp{
\textbf{\ding{182} Patch path selection.} We first divide patch anchors into sub-graphs based on their connectivity and enumerate all the simple paths from the entry to exit nodes in each sub-graph as candidate patch paths, i.e., $\textbf{D2} \to \textbf{C3}$, $\textbf{G1} \to \textbf{F2}$, and $\textbf{C3}$. Subsequently, we aim to select the most distinguishable and representative by establishing two priorities.
\begin{itemize}[leftmargin=.25in]
  \item \emph{Pry-1: Exclusive to a single reference function.} The patch path includes anchors that are present in one reference function and are absent in the other.
  \item \emph{Pry-2: Higher weight.} The patch path includes anchors with higher weights. 
\end{itemize}
Therefore, $\textbf{D2} \to \textbf{C3}$ (The only candidate) in~\Cref{fig: anchor path extraction}a and $\textbf{G1} \to \textbf{F2}$ (satisfy \textbf{Pry-1} and \textbf{Pry-2}) in~\Cref{fig: anchor path extraction}b are selected as patch paths, denoted as $AP_{patch}^{ref}$.
}

\cp{
\textbf{\ding{183} Context path extraction.} We further extract the context paths by exploring both backward and forward directions from the patch path to differentiate the patch-similar code blocks, as discussed in~\Cref{sec: motivating example}, from the real patch code blocks. When extracting the backward context path, we start from the first anchor of $AP_{patch}$ and end the search at the entry node of AG with the greedy strategy~\cite{greedy}. This process involves recursively selecting the predecessor anchor with the highest weight from the current anchor and recording it as part of the backward path. The extraction of the forward context path follows similar steps by starting from the last anchor of $AP_{patch}$ and ending at the exit node of AG. The context path comprises both the backward and forward paths, denoted as $AP_{bw}^{ref}$ and $AP_{fw}^{ref}$
Take the backward path extraction in~\Cref{fig: anchor path extraction}b as an example, starting from the patch anchor \textbf{G1}, we examine the predecessors of \textbf{G1} and find that \textbf{B1} has a higher weight than \textbf{C1}. Thus, $\textbf{A1} \to \textbf{B1}$ is extracted as the backward path of the patch path $\textbf{G1} \to \textbf{F2}$. 
}

\textbf{Signature Definition.}
Based on the extracted patch path and context path (backward and forward), we present the definition of our signature as follows:
\begin{equation}\label{eq: sig}
sig = (AP_{patch}^{ref}, AP_{bw}^{ref}, AP_{fw}^{ref})
\end{equation}
There are two different signatures derived from $f_{vul}^{ref}$ and $f_{fix}^{ref}$, denoted as $sig_{vul}$ and $sig_{fix}$, respectively. 

\subsection{Patch Detector}
The patch detector is responsible for filtering the irrelevant functions and locating the patch code in the target function (\tgt{}) with the given signatures. For \tgt{}, we generate its anchor graph ($AG_{tgt}$) in the same way described in~\Cref{sec: anchor graph construction} for detection.

\subsubsection{Irrelevant Function Filtering}\label{sec: irrelevant function filtering}
Before locating the patch code, we aim to filter out irrelevant functions by evaluating all the anchors in $AG_{tgt}$ at a coarse-grained level. To achieve this, we compute the matching proportion of unique anchor values between the AG of reference function $AG_{ref}$ and  $AG_{tgt}$ using the equation $\rho = \frac{m}{N}$, where $m$ represents the number of matched unique anchor values between the two AGs, and $N$ denotes the total number of unique anchor values in $AG_{ref}$. Suppose neither the proportion between the vulnerable function and target function ($\rho_{(vul, tgt)}$) nor the proportion between the fixed function and target function ($\rho_{(fix, tgt)}$) exceeds the threshold $T_{iff}$. In that case, the target function will be classified as irrelevant without further matching. 

\subsubsection{Anchor Path Matching}\label{sec: anchor path matching}
\cp{Given a reference anchor path $AP_{ref}$, the process of anchor path matching aims to identify the most similar anchor path within $AG_{tgt}$. 
Two key challenges must be addressed during this process. 
\emph{First, the number of false positives generated by the BCSD tool increases with the growth in the number of candidate functions.} Given that a binary typically contains thousands of functions, directly matching the invoked functions of the reference function with all functions in the target binary to detect anchors of the ``CALL'' type is impractical.
\emph{Second, in an $AG_{tgt}$ with multiple nodes, the number of candidate anchor paths for matching becomes substantial.} For example, if each anchor in $AP_{ref}$ has 4 candidates from $AG_{tgt}$ and the length of $AP_{ref}$ is 5, the total number of candidate anchor paths amounts to $4^5 = 1024$. Therefore, enumerating all candidate anchor paths in $AG_{tgt}$ and matching them with $AP_{src}$ becomes time-consuming and impractical.
}
To solve the challenges, we decide to reduce the number of candidate functions and anchor paths for matching from two perspectives: 

\cp{\emph{1) Narrow down the search space based on values and caller constraints.}
We select candidate anchor set $S_{cand}$ for each anchor of $AP_{ref}$ from $AG_{tgt}$ by matching anchors with the same value and having more matched auxiliary information. For anchors of the "CALL" type, we first attempt to match based on the function name, if available. Otherwise, We employ the BCSD tool to match "CALL"-type anchors by comparing the invoked functions of the reference exclusively with those of \tgt{}, as other functions are not invoked by \tgt{} and therefore need not be considered.
For example, consider a target binary containing $N$ functions ($f_1, f_2, \ldots, f_N$), where the target function \tgt{} invokes only three functions ($f_1, f_2, f_3$). When matching the "CALL" anchor $f_k$ invoked by the reference function, we compare $f_k$ exclusively with ($f_1, f_2, f_3$) instead of all $N$ functions. The similarity between two functions must exceed $T_{bcsd}$; otherwise, the invoked function is deemed unmatched.
As a result, we derive the list of candidate anchor sets for $AP_{ref}$, represented as $L_{S_{cand}} = \langle S_{cand}^1, S_{cand}^2, \ldots, S_{cand}^n \rangle$, where $S_{cand}^i$ denotes the candidate anchor set corresponding to $ac_i$ in $AP_{ref}$.
}

\emph{2) Filter out false positives based on the distance between adjacent anchors in $AP_{ref}$}.
Not all candidate anchors should be considered for matching; for example, candidates that are too distant from the previous one or are unreachable should be disregarded. Thus, we iterate through all the candidate sets, selecting one anchor from each to form the desired paths, while leveraging the distance between anchors in $AP_{ref}$ to narrow down the set of potential anchors. 
The distance between the two anchors in AG is defined as follows:
\begin{equation}\label{eq: ac distance}
    d(ac_1, ac_2) = \begin{cases}
        Infinite &   \text{There is no path from $ac_1$ to $ac_2$}, i.e., unreachable \\
       \ell(p) & \text{The length of the shortest path p from $ac_1$ to $ac_2$} 
    \end{cases}
\end{equation}
Suppose the currently selected anchor from $S_{cand}^i$ is $ac_{cur}$; in that case, we will select only those anchors in $S_{cand}^{i+1}$ that are closest to $ac_{cur}$ (i.e., minimum distance) and ignore other candidates in $S_{cand}^{i+1}$ directly.

\begin{algorithm}[t]
    \caption{Anchor path matching}
    \label{alg: anchor path matching}
    \LinesNumbered
    \SetAlgoLined
    \SetKwFunction{PathMatch}{PathMatch}  
    \SetKwFunction{length}{length}  

    \KwIn{Reference anchor path $AP_{ref}$ and the corresponding list of candidate sets $L_{S_{cand}}$}
    \KwOut{A list of matched anchor paths $L_{AP_{match}}$}

    $L_{AP_{match}}, max\_len \gets \langle\rangle, 0$\;
    \PathMatch($0, \langle\rangle$)\;
    \textbf{return} $L_{AP_{match}}$\; 
    \SetKwProg{Fn}{Function}{:}{}
    \Fn{\PathMatch{$index, AP$}}{
     
        \uIf(\tcp*[h]{The matching has reached the end}){$index == \length(AP_{ref})$}{ 
            \uIf{$   
            \length(AP_{match}) > max\_len$}{    
                $L_{AP_{match}}, max\_len \gets \langle AP \rangle, \length(AP_{match})$\;    
            }
            \uElseIf{$\length(AP_{match}) == max\_len$}{
                $L_{AP_{match}}$.add($AP_{match}$)\;
            }
            \textbf{return}\;
        }
        \uIf(\tcp*[h]{The start of matching}){$\length(AP_{match}) == 0$}{
            $S_{cand} \gets L_{S_{cand}}[index]$\; 
        }
        \uElse{
            $S_{cand} \gets$ Choose candidate anchors closest to $AP_{match}[-1]$ from $L_{S_{cand}}[index]$\;
        }
        \uIf(\tcp*[h]{No candidates detected for $AP_{ref}[index]$}){$\length(S_{cand}) == 0$}{
            \PathMatch($index + 1, AP_{match}$)\;
        }
        \uElse{
            \For{$ac \in S_{cand}$}{ 
                \PathMatch($index + 1, AP_{match} + ac$)\; 
            }
        }

    }

\end{algorithm}

Based on the above strategies, we propose the anchor path matching algorithm~\ref{alg: anchor path matching}. 
From the beginning, $L_{AP}$ and $max\_len$ are initialized as an empty list and 0, which are used to store the matched anchor paths and the current maximum length of matched anchor path $AP_{match}$, respectively (line 1). 
The core matching process is carried out using a depth-first search algorithm~\cite{depthFirstSearch}, implemented in the function \texttt{PathMatch} (lines 4 to 20). The function accepts two parameters: $index$ (the current index of the anchor being matched in $AP_{ref}$) and $AP_{match}$ (the current matched anchor path). 
In \texttt{PathMatch}, the first step is to verify if the matching has reached the end by comparing $index$ to the length of $AP_{ref}$. If yes, $L_{AP_{match}}$ and $max\_len$ are updated as shown in lines 6 to 9. Otherwise, the candidate set $S_{cand}$ is selected, and $AP_{match}$ is updated for the next matching round by checking if $AP_{match}$ and $S_{cand}$ are empty, as described in lines 11 to 20.

For the context path, we require that the matching ratio surpasses the threshold $T_{cpm}$ to eliminate incorrectly identified paths while allowing minor discrepancies in the context. For the patch path, we require the anchors to be fully matched, ensuring the patch's presence.

\subsubsection{Patch Path Verification}
As outlined in~\Cref{sec: motivating example}, patch-similar code blocks can mislead patch code localization when relying solely on the patch path. Therefore, it is essential to incorporate the context to verify the authenticity of the patch path.
\emph{Our insight is that patch-similar code blocks fail to uphold the control flow relationship between themselves and the context of the patch.}
Specifically, we define the distance between two anchor paths \(d(AP_a, AP_b)\), \(AP_{a} = \langle a_1, a_2, \ldots, a_n \rangle\) and \(AP_{b} = \langle b_1, b_2, \ldots, b_m \rangle\), as the distance between the last anchor in \(AP_a\) and the first anchor in \(AP_b\) (i.e., \(d(a_n, b_1)\)). 
The matched anchor paths \(AP_{patch}^{match}\) and  \(AP_{bw}^{match}\), \(AP_{fw}^{match}\) in the previous section~\Cref{sec: anchor path matching} are only considered validated if they satisfy the following condition:
\begin{equation}
    d(AP_{bw}^{match}, AP_{patch}^{match}) \le d(AP_{bw}^{ref}, Ap_{patch}^{ref}) \land d(AP_{patch}^{match}, AP_{fw}^{match}) \le d(AP_{patch}^{ref}, AP_{fw}^{ref})
\end{equation}
Additionally, if multiple patch paths are validated, we select the one with the minimum distance to the context paths.

\subsection{Function Classifier}
Using the located patch, context paths, and the signatures of two reference functions, the function classifier seeks to categorize \tgt{} as vulnerable, fixed, or irrelevant through a decision tree.

\begin{figure}[ht]
  \centering
  \includegraphics[width=\linewidth]{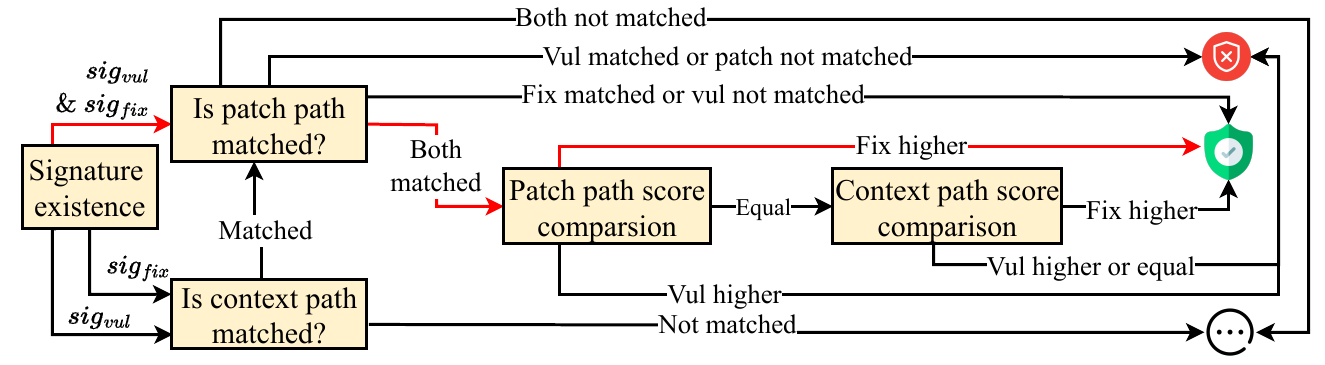}
  \caption{The decision tree of function classifier. The red line is highlighted as an example.}
  \label{fig: category decision tree}
\end{figure}

In our design, a vulnerable function must include the patch path in $sig_{vul}$ or the context path in $sig_{fix}$ when $sig_{vul}$ is \texttt{NULL}. Likewise, a fixed function must contain the patch path in $sig_{fix}$ or the context path in $sig_{vul}$ when $sig_{fix}$ is \texttt{NULL}. Functions that do not meet the requirements should be identified as irrelevant. According to these principles, we construct the decision tree with five related decision points as shown in~\Cref{fig: category decision tree} to classify the functions. 
The scores for the matched anchor path is calculated as shown in~\Cref{eq: path score}.
\begin{equation}
s = \frac{\sum_{i=0..|AP_{ref}|} LCS(AP_{ref}[i].aux, AP_{match}[i].aux)}{|AP_{ref}|}
\label{eq: path score}
\end{equation}
Here, $AP_{ref}$ and $AP_{match}$ denote the reference anchor path and the matched anchor path, respectively. By enumerating the anchors in $AP_{ref}$ and $AP_{match}$, we compare the corresponding auxiliary information in each anchor using the longest common sequence (LCS)~\cite{LCS}. The average matching score of each anchor in $AP_{ref}$ is computed as the final matching score of a reference anchor path. The context score is the sum of the backward path score and forward path score.

To explain the whole process, we take the decision route in the red line as an example. In this route, two signatures both exist, leading to the verification of patch paths. Since the two patch paths are both matched, we further calculate the patch score between the matched patch path and the reference patch path in the two signatures and find that the patch path in $sig_{fix}$ gets a higher value than the patch path in $sig_{vul}$, resulting in the final classification of \tgt{} as fixed. 

\section{Evaluation}
We aim to address the following \textbf{research questions (RQs)}.

\begin{itemize}[leftmargin=*]
    \item \textbf{RQ1.} How accurate is \toolname{} in identifying vulnerable and fixed functions with the same compilation settings, across different optimization levels, and different compilers, compared to the state-of-the-art approaches?
    \item \textbf{RQ2.} How does \toolname{}'s efficiency in the detection process compare to state-of-the-art approaches?
    \item \textbf{RQ3.} How effective is \toolname{} when irrelevant functions are involved, in comparison to state-of-the-art approaches? To what extent do the two components of \toolname{}, namely irrelevant function filtering and patch path verification, contribute to enhancing accuracy (i.e., through an ablation study)?
\end{itemize}

\textbf{RQ1} and \textbf{RQ2} are employed to compare \toolname{} with existing patch presence test approaches, using the same experimental setup as in their studies~\cite{Xu2020PatchBV, Yang2023TowardsPB}. \textbf{RQ3} is used to compare \toolname{} with baselines in a more challenging and practical scenario, as described in~\Cref{srvd task}, where irrelevant functions are present.

\begin{figure}[ht]
  \centering
  \includegraphics[width=\linewidth]{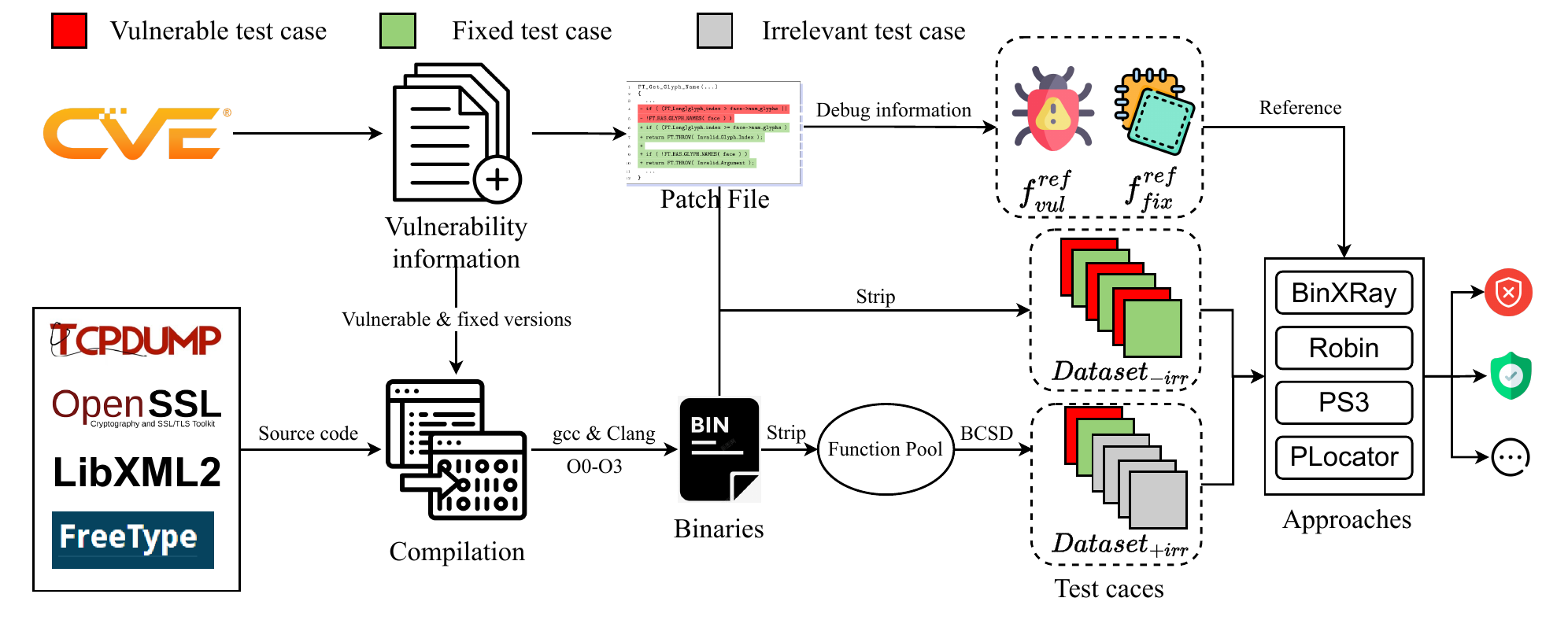}
  \caption{The process of dataset construction and vulnerability detection with different approaches.}
  \label{fig: dataset construction}
\end{figure}

\subsection{Experimental Setup}\label{sec: experimental setup}
\subsubsection{Dataset}\label{sec: dataset}
~\Cref{fig: dataset construction} presents the pipeline of how we generate our dataset for testing. To evaluate our approach, we select four real-world well-known projects from various application aspects for evaluation, which have also been well analyzed by previous works~\cite{Zhan2023PS3PP, Yang2023TowardsPB}. These four projects involved protocol encryption, packet, XML analyzer, and font render. Then, we collect the vulnerability information of these projects (e.g., vulnerable and fixed versions, patch links) from NVD~\cite{NVD} and compile the corresponding vulnerable and fixed binaries with two compilers (gcc v9.4.0 and Clang 6.0) and four different optimization levels (O0 to O3). 
The first seven columns in ~\Cref{tab:dataset_composition} show the statistic results of vulnerability and project information. In total, we have obtained 73 CVEs with 224 source vulnerable and fixed functions (i.e., $f_{vul}^{src}$ and $f_{fix}^{src}$). The 73 CVEs encompass seven prevalent vulnerability types (e.g., memory corruption, overflow, etc.), with the size of the vulnerable functions (measured by the number of basic blocks) ranging from 1 to 643. 
Ultimately, We extract features of each approach using binaries with \emph{gcc compiler}, \emph{O0} optimization level and strip the binaries for testing dataset construction.

\begin{table}[]
\caption{Statistics of our dataset. $f_{vul}^{src}$ and $f_{fix}^{src}$ refer to the source vulnerable and fixed functions, respectively. $f_{vul}, f_{fix}, f_{irr}$ refer to the vulnerable, fixed, and irrelevant functions in binaries, respectively. $\overline{f_{bin}}$ is the average number of functions in binaries.}
\label{tab:dataset_composition}
\begin{tabular}{crrrrrr|rr|rrr}
\hline \hline
\multicolumn{7}{c|}{\textbf{Original Data information}} & \multicolumn{2}{c|}{\textbf{$Dataset_{-irr}$}} & \multicolumn{3}{c}{\textbf{$Dataset_{+irr}$}} \\ \hline 
Projects &
  \multicolumn{1}{l}{\# CVE} &
  \multicolumn{1}{l}{\# $f_{vul}^{src}$} &
  \multicolumn{1}{l}{\# $f_{fix}^{src}$} &
  \multicolumn{1}{l}{\# Version} &
  \multicolumn{1}{l}{\# Binary} &
  \multicolumn{1}{l|}{\# $\overline{f_{bin}}$} &
  \multicolumn{1}{c}{\# $f_{vul}$} &
  \multicolumn{1}{c|}{\# $f_{fix}$} &
  \multicolumn{1}{c}{\# $f_{vul}$} &
  \multicolumn{1}{c}{\# $f_{fix}$} &
  \multicolumn{1}{c}{\# $f_{irr}$} \\ \hline
OpenSSL    & 34   & 50    & 50    & 21  & 105  & 5,943  & 248                    & 248                   & 311          & 311          & 10,091          \\
Freetype   & 7    & 9     & 9     & 5   & 25   & 1,305  & 45                     & 45                    & 45           & 39           & 1,822           \\
Tcpdump    & 25   & 39    & 39    & 2   & 10   & 1,096  & 182                    & 182                   & 172          & 164          & 11,088          \\
Libxml2    & 7    & 14    & 14    & 4   & 20   & 3,306  & 70                     & 70                    & 76           & 41           & 3,090           \\ \hline
Total      & 73   & 112   & 112   & 32  & 160  & -      & 545                    & 545                   & 604          & 555          & 26,091          \\ \hline \hline
\end{tabular}
\end{table}

To address the research questions, we construct two datasets to compare \toolname{} with baselines, both with and without interference from irrelevant functions. Each test case represents a target binary function evaluated using different approaches.

\begin{itemize}[leftmargin=.25in]
\item $Dataset_{-irr}$ (\emph{Dataset without irrelevant functions).} This dataset only consists of the vulnerable and fixed functions, which is designed for \textbf{RQ1} and \textbf{RQ2}. In total, there are 1,090 test cases as shown in columns 8 to 9 in~\Cref{tab:dataset_composition}.

\item $Dataset_{+irr}$ (\emph{Dataset with irrelevant functions).} Except for all the vulnerable and fixed functions in $Dataset_{-irr}$, we also randomly select irrelevant functions from the binaries to form this dataset, which is designed for \textbf{RQ3}. Given that the maximum average number of functions in binaries is 5,943 (column 7 in~\Cref{tab:dataset_composition}), we set the function pool size to 6,000. 
Subsequently, we set the vulnerable functions as the query function and utilize BCSD tool jTrans~\cite{Wang2022jTransJT} to recall the top-50 functions (ensure the vulnerable functions being recalled) by following the process of 1-day vulnerability detection as described in~\Cref{fig: vul_search_task} (i.e., potentially vulnerable function set). In total, there are 27,250 test cases, as displayed in columns 10 to 12 of~\Cref{tab:dataset_composition}. 
\end{itemize}

\cp{To determine the thresholds mentioned in our study, we randomly selected 20\% of the test cases (i.e., 111 vulnerable, 108 fixed, and 4,381 irrelevant) from $Dataset_{+irr}$ for threshold selection. The remaining test cases in $Dataset_{-irr}$ (i.e., 453 vulnerable and 453 fixed) and $Dataset_{+irr}$ (i.e., 493 vulnerable, 447 fixed, and 21,710 irrelevant) were used to evaluate the effectiveness of the approaches.}

\subsubsection{Experiments and Metrics}
\cp{To compare the effectiveness of the approaches on different compilers and optimizations, we set three experiments: \emph{1) Same}. The reference and target functions are from binaries with the same compiler (gcc) and optimizations level (O0). \emph{2) XO (Cross-optimizations)}. The reference and target functions are from binaries with different optimization levels (O1 to O3). \emph{3) XC (Cross-compilers)}. The reference and target functions are from binaries with different compilers (gcc and Clang). 
Same to~\Cref{sec: preliminary study on BCSD}, we use TPR and FPR as the metrics to evaluate the effectiveness of the approaches. 
During our experiments, we observed that the baselines supported only a subset of the test cases in our dataset. The primary causes of this limitation were signature generation failures, timeouts, and unforeseen exceptions. To facilitate a fair comparison, we introduced three additional metrics: support rate $SR=\frac{TC_{s}}{TC_{all}}$ ($TC_{s}$ is the number of support test cases and $TC_{all}$ is the number of all test cases), $TPR_{s}$ ($TPR$ on supported test cases), and $FPR_{s}$ ($FPR$ on supported test cases).
}

\subsection{Baseline Methods}
We compare \toolname{} with three state-of-the-art patch presence test approaches, one is syntactic-based while the other two are semantic-based methods. 
\begin{itemize}[leftmargin=.25in]
\item \textbf{BinXray~\cite{Xu2020PatchBV}}. A syntactic-based patch presence test approach, which uses basic block mapping to extract the execution traces of the patch from two reference binary functions and compare them to the trace extracted from the target function.
\item \textbf{Robin~\cite{Yang2023TowardsPB}}. A semantic-based patch presence test approach.
It employs symbolic execution to extract the malicious function input (MFI) that triggers the vulnerable code. The MFI is then fed into the target, and similarities are calculated with the captured semantic features to determine the presence of a patch.
\item \textbf{PS3~\cite{Zhan2023PS3PP}}. A semantic-based approach extracts the features from the vulnerable and fixed functions through symbolic execution and matches them in the target binary function.
\end{itemize}

For the above baselines, we use the original implementations as described in their papers. We regard the target functions as \texttt{``irrelevant''} when the baselines return an uncertain result (e.g., with a score of $0$) or fail to identify the target (e.g., "unknown" or time out).
We exclude other related approaches from the experiments for two main reasons: 1) They did not release the source code (e.g., PDiff~\cite{Jiang2020PDiffSP}, SPAIN~\cite{Xu2017SPAINSP}). 2) They perform worse than our select baselines in the previous studies (e.g., Fiber~\cite{Zhang2018PreciseAA}, PMatch~\cite{Lang2021PMatchSP})
To evaluate the effects of two components (i.e., irrelevant function filtering and patch path verification) in \toolname{}, we further set up three configurations:

\begin{itemize}[leftmargin=.25in]
    \item \toolname{}$_{-IFF}$: \toolname{} without \underline{I}rrelevant \underline{F}unction \underline{F}iltering.
    \item \toolname{}$_{-PPV}$: \toolname{} without \underline{P}atch \underline{P}ath \underline{V}erification.
    \item \toolname{}$_{-Both}$: \toolname{} without the both components.
\end{itemize}

\subsection{Implementation}
\subsubsection{Environment and Tools.}
\toolname{} is implemented in Python with 3,365 lines of code. We utilize IDA Pro 7.5~\cite{ida} and IDAPython to disassemble the binary functions and construct the CFGs. 
All experiments, except for BCSD, were conducted on a laptop with an Intel 16-Core i9-9880 2.30GHz processor, 64 GB of memory, running Ubuntu 22.04 OS. We implemented jTrans on a server equipped with an Intel Xeon Gold 5218 CPU @ 2.30GHz, 1 TB of memory, and 2 Nvidia Tesla V100 GPUs (32GB each). 

\begin{figure}[t]
  \centering
  \includegraphics[width=\linewidth]{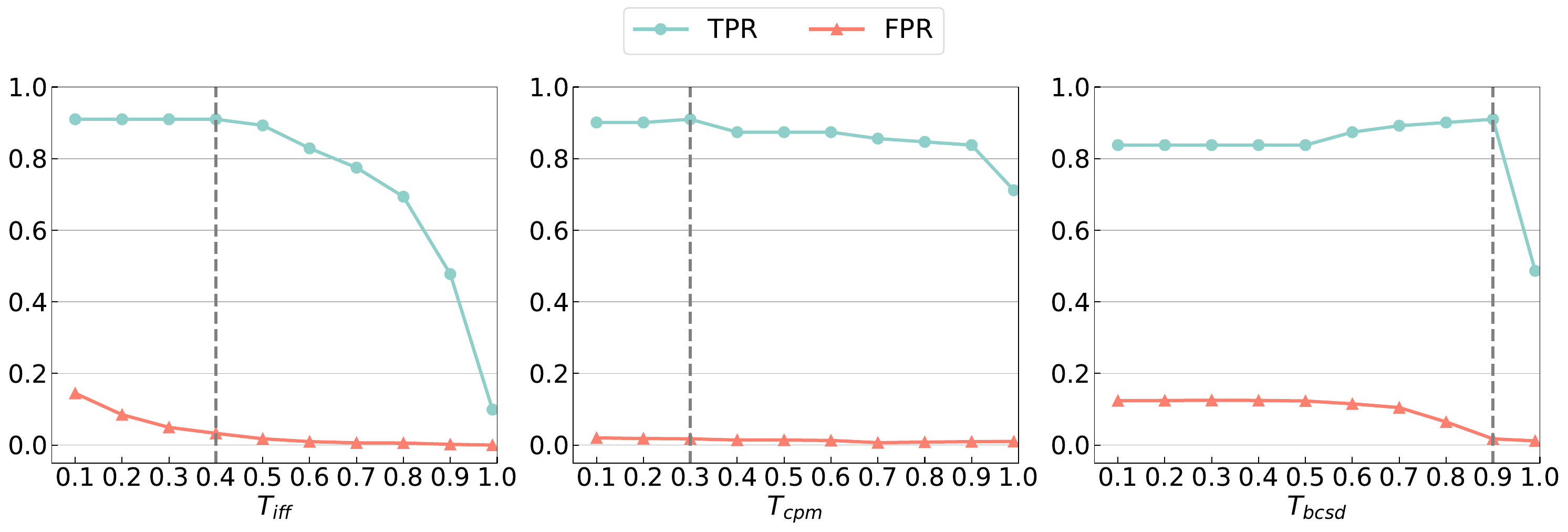}
  \caption{Threshold selection. The X-axis is the value of thresholds and the Y-axis is the value of metrics. The best thresholds are highlighted with grey dotted lines.}
  \label{fig: threshold selection}
\end{figure}

\subsubsection{Threshold Selection.}
We vary the three thresholds $T_{iff}$ (irrelevant function filtering), $T_{cpm}$ (patch path verification), and $T_{bcsd}$ (invoked functions BCSD) from $0.1$ to $1$ to identify the thresholds yielding the best results (i.e., highest TPR and lower FPR). The threshold selection results are shown in~\Cref{fig: threshold selection}, in which $T_{iff}$, $T_{cpm}$ and $T_{bcsd}$ yield the best results at $0.4$, $0.3$ and $0.9$, respectively.

\subsection{RQ1: Result of Vulnerability Detection without Irrelevant Functions}
In this research question, we focus solely on identifying vulnerable functions among fixed functions, consistent with previous works. For each test case in $Dataset_{-irr}$, we categorize it based on each method's output and compute the metrics presented in~\Cref{tab: rq1_res}. 

\cp{\paragraph{Syntactic-based method (BinXRay) result analysis}
BinXRay supports only 58.4\%, 18.9\% and 10.8\% of test cases in $Dataset_{-irr}$ for the three experiments, respectively. 
Upon examining the BinXRay source code, we found that it struggles to track traces or distinguish code differences when the disparity between the target and reference is substantial, primarily due to the failure of its block mapping algorithm. Therefore, BinXRay is entirely ineffective when the source and target binaries are compiled using different compilers or optimizations since these two compilation settings alter the binary code syntax much. 
}

\begin{table}[]
\caption{The results on $Dataset_{-irr}$ (\%).  }
\label{tab: rq1_res}
\begin{adjustbox}{max width=\textwidth}
\begin{threeparttable}
\begin{tabular}{c|rrrrr|rrrrr|rrrrr|rr}
\hline
\hline
\textbf{Experiment} & \multicolumn{5}{c|}{\textbf{Same}} & \multicolumn{5}{c|}{\textbf{XO}} & \multicolumn{5}{c|}{\textbf{XC}} & \multicolumn{2}{c}{\textbf{Average}} \\ \hline
\textbf{Metric} &
  \multicolumn{1}{c}{$TPR_{s}$} &
  \multicolumn{1}{c}{$FPR_{s}$} &
  \multicolumn{1}{c}{$SR$} &
  \multicolumn{1}{c}{$TPR$} &
  \multicolumn{1}{c|}{$FPR$} &
  \multicolumn{1}{c}{$TPR_{s}$} &
  \multicolumn{1}{c}{$FPR_{s}$} &
  \multicolumn{1}{c}{$SR$} &
  \multicolumn{1}{c}{$TPR$} &
  \multicolumn{1}{c|}{$FPR$} &
  \multicolumn{1}{c}{$TPR_{s}$} &
  \multicolumn{1}{c}{$FPR_{s}$} &
  \multicolumn{1}{c}{$SR$} &
  \multicolumn{1}{c}{$TPR$} &
  \multicolumn{1}{c|}{$FPR$} &
  \multicolumn{1}{c}{$TPR$} &
  \multicolumn{1}{c}{$FPR$} \\ \hline
BinXray             & 92.3  & 5.8   & 58.4 & 53.9 & 3.4  & 71.4 & 23.1 & 18.9 & 13.1 & 4.5  & 36.4 & 20.0 & 10.8 & 4.1  & 2.1  & 23.7              & 3.3              \\
Robin               & 95.1  & 30.5  & 92.1 & 87.6 & 28.1 & 82.5 & 63.4 & 92.1 & 76.0 & 58.4 & 52.9 & 33.7 & 89.2 & 47.4 & 29.9 & 70.4              & 38.8             \\
PS3                 & 91.7  & 19.4  & 80.9 & 74.2 & 15.7 & 69.9 & 56.5 & 80.9 & 56.6 & 45.7 & 61.8 & 50.0 & 78.4 & 48.5 & 39.2 & 59.7              & 33.5             \\ \hline
\rowcolor[HTML]{C0C0C0} 
\textbf{PLocator} &
  \textbf{93.3} &
  \textbf{12.4} &
  \textbf{100.0} &
  \cellcolor[HTML]{C0C0C0}\textbf{93.3} &
  \cellcolor[HTML]{C0C0C0}\textbf{12.4} &
  \textbf{85.8} &
  \textbf{22.5} &
  \textbf{100.0} &
  \cellcolor[HTML]{C0C0C0}\textbf{85.8} &
  \cellcolor[HTML]{C0C0C0}\textbf{22.5} &
  \textbf{84.5} &
  \textbf{13.4} &
  \textbf{100.0} &
  \cellcolor[HTML]{C0C0C0}\textbf{84.5} &
  \cellcolor[HTML]{C0C0C0}\textbf{13.4} &
  \textbf{87.9} &
  \textbf{16.1} \\ \hline
  \hline
\end{tabular}%

\begin{tablenotes}
      \item[1] The low FPR of BinXRay is due to the fact that it returns ``unknown'' for most test cases, which are classified as irrelevant by default.
\end{tablenotes}
\end{threeparttable}
\end{adjustbox}

\end{table}

\cp{\paragraph{Semantic-based methods (Robin and PS3) results analysis}
Robin and PS3 demonstrate relatively stable performance compared to BinXRay, particularly in the \emph{XO} and \emph{XC} experiments. However, compared to the \emph{Same} experiment, they retrieved an average of 29.6\% and 27.8\% fewer vulnerable functions in the \emph{XO} and \emph{XC} experiments, respectively, while increasing the FPR by 57.2\% and 169.8\%. 
}

\begin{figure}[ht]
  \centering
  \includegraphics[width=\textwidth]{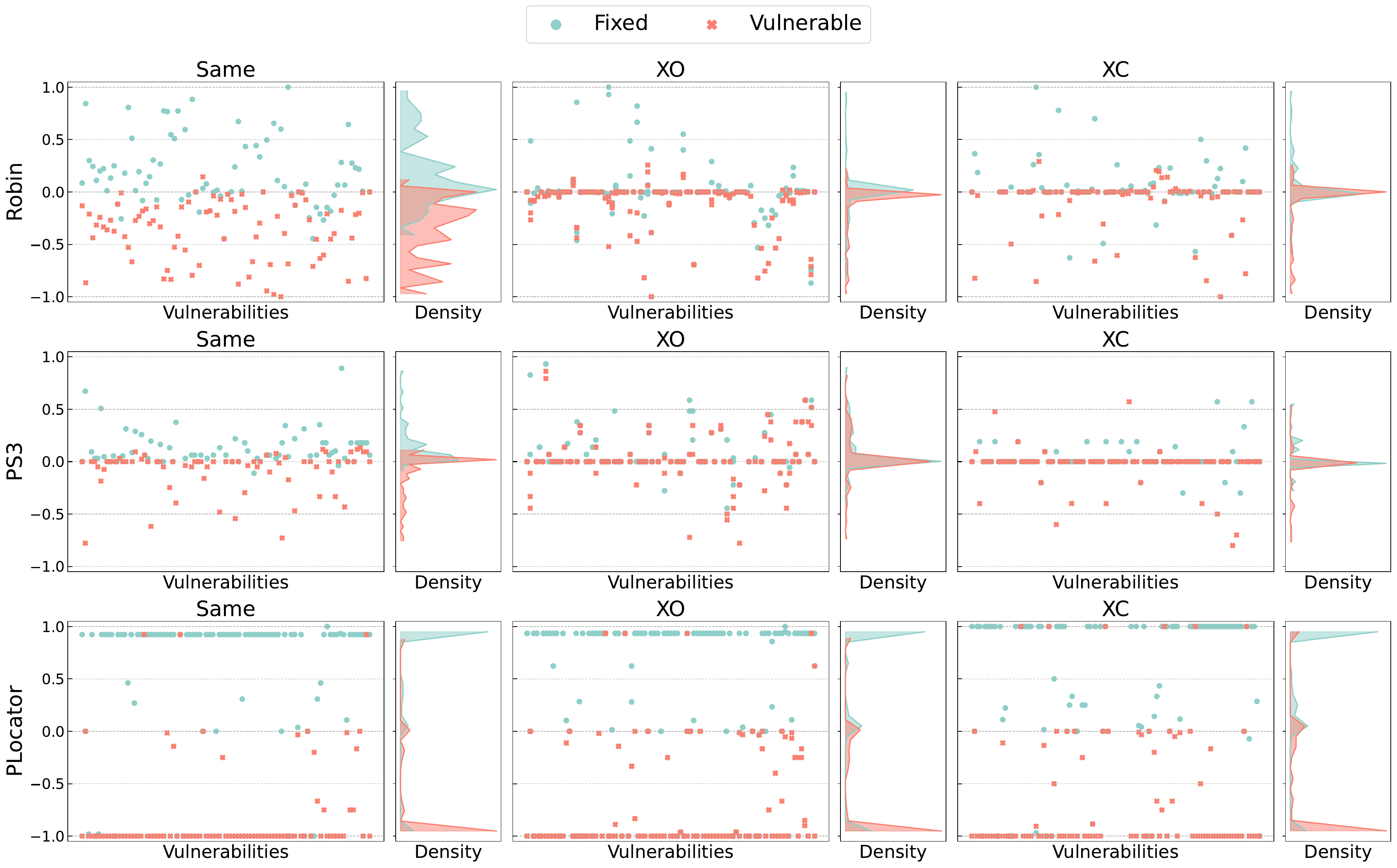}
  \caption{Test case score scatter plot and curve plot of Robin PS3, and PLocator on $Dataset_{-irr}$. The legend indicates the true category of the test case. The \emph{y-axis} refers to the score, and the \emph{x-axis} in scatter and curve refers to the different vulnerabilities and probability density of test cases, respectively.}
  \label{fig: score dist rq1}
\end{figure}

\cp{To clearly illustrate their predictions for different test cases, we plot the score of each test case for three approaches as shown in~\Cref{fig: score dist rq1}. 
All three approaches classify the target functions based on the matched features of two reference functions (i.e., vulnerable and fixed). Thus, we calculate the score by subtracting the score of matched vulnerable features from the score of matched fixed features and then normalizing it to the range $(-1, 1)$.  
For each experiment under a specific approach (e.g., (Robin, Same)), two figures are provided:  
\begin{itemize}[leftmargin=.25in]
    \item \emph{Test Case Score Scatter Plot (left figure).} Each test case is plotted as a node based on the reference vulnerability and its score predicted by the approach. The \emph{x-axis} represents different vulnerabilities, while the \emph{y-axis} indicates the score of test cases under the corresponding vulnerabilities.
    \item \emph{Test Case Score Curve Plot (right figure).} The \emph{x-axis} represents the probability density of test cases at a specific score, sharing the \emph{y-axis} with the scatter plot. A higher \emph{x-axis} value indicates a greater number of test cases with the corresponding score.
\end{itemize}
\emph{Under ideal circumstances, all vulnerable test cases should yield negative scores, whereas fixed ones should yield positive scores.}
As shown in the figure, Robin and PS3 accurately predict the test cases in the \emph{Same} experiments but generate more false positives and false negatives in the other two experiments. Moreover, the absolute value of scores decreased from the \emph{Same} experiment to the other two experiments. 
\emph{In that case, although they claim to select semantic features resilient to varying compilation settings, many of the features they depend on are lost under different compiler and optimization conditions, causing the boundary between vulnerable and fixed functions to become indistinct.}  
The results of PS3 are similar to Robin's since they both depend on emulation of the target function and match features from the reference against the target. The key distinction between them lies in their choice of semantic features and the collected traces from the target function. 
For feature selection, PS3 assumes that function names remain unchanged in the target function and uses function call names as features. In contrast, Robin selects function call arguments and ignores function call names, achieving better performance on datasets containing stripped binaries. 
For trace collection, PS3 enumerates all blocks of the target function. In contrast, Robin focuses only on traces related to the vulnerable and fixed code, based on their Malicious Function Input (MFI), which mitigates interference from irrelevant code to some extent.
}

\paragraph{PLocator result analysis.}
Across all the experiments, \toolname{} significantly outperformed the baselines, achieving an average improvement of 24.9\% in $TPR$ and a reduction of 52.1\% in $FPR$ compared to the second-best method. As depicted in~\Cref{fig: score dist rq1}, \toolname{} effectively separates the test cases into their true categories, even under different compilers and optimizations. The similarity in the absolute score values across the three experiments highlights the robustness of the features selected by \toolname{}. To further prove the stability of our selected features, we extracted all key instructions (anchors) from the test case in $Dataset_{-irr}$ and mapped their addresses to source code line numbers using debug information. The results show that 91\% anchors persist across compilers and optimizations.

\begin{figure}[htbp]
    \centering
    \begin{subfigure}{0.32\textwidth}
        \centering
        \includegraphics[width=\linewidth]{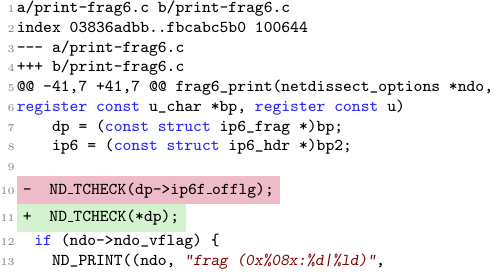}
        \caption{Patch of CVE-2017-13031~\cite{cve-2017-13031}.}
        \label{fig: patch_cve_2017_13031}
    \end{subfigure}
    \hfill
    \begin{subfigure}{0.32\textwidth}
        \centering
        \includegraphics[width=\linewidth]{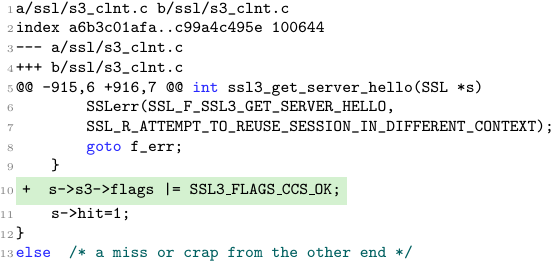}
        \caption{Patch of CVE-2014-0224~\cite{cve-2014-0224}}
        \label{fig: patch_cve_2014_0224}
    \end{subfigure}
    \hfill
    \begin{subfigure}{0.32\textwidth}
        \centering
        \includegraphics[width=\linewidth]{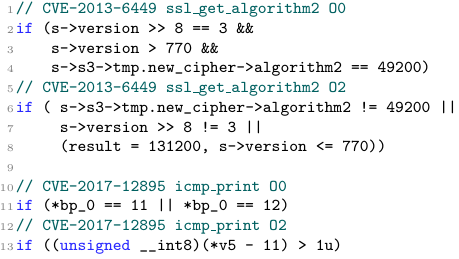}
        \caption{Changes caused by optimization}
        \label{fig: anchor_path_change_by_optim}
    \end{subfigure}
    \caption{Examples of \toolname{}'s test cases}
    \label{fig: test_cases_example}
\end{figure}

We manually analyze \toolname{}'s false positives and false negatives to conclude three primary causes:

\emph{1) undetectable patch modifications.} Some patches will not affect the anchors extracted from the two references, resulting in the same signatures generated for vulnerable and fixed functions. For example, the patch of \texttt{CVE-2014-0224} only added an assignment statement as shown in~\Cref{fig: patch_cve_2014_0224}, resulting in the signature of vulnerable and fixed to be the same.

\emph{2) Anchor path variation caused by compilers and optimizations.} In some special cases, the anchor paths across different compilers and optimizations may show disparity. For example, as illustrated in~\Cref{fig: anchor_path_change_by_optim}, the order of three conditions in the function \texttt{ssl\_get\_algorithm2} at O0 changes when the function is compiled with O2, leading to a corresponding change in the order of anchors in the patch path. The two conditions in the function \texttt{icmp\_print} at O0 are optimized as a single condition by casting the expression \texttt{*v5 - 11} as an unsigned integer, which includes the same semantics. 

\emph{3) Insufficient patch or context information.}
\toolname{} fails to detect the patch and context path precisely due to insufficient information, such as an anchor path with only a few anchors or does not contain rich auxiliary information. These anchor paths can be mistakenly identified in other similar code blocks and result in false identification.

\paragraph{\toolname{} for tiny patch modification identification.}
Owing to the patch code localization and decision tree of the function classifier, \toolname{} can even detect patches with tiny modifications. For instance, the patch for \texttt{CVE-2017-13031} modifies only a single parameter, as illustrated in~\Cref{fig: patch_cve_2017_13031}. \toolname{} successfully located the patch code in the target and accurately classified the true label by determining which parameter (i.e., represented as auxiliary information) is closer to the target.    

\cp{\paragraph{Patch similar code interference analysis} Based on the statistics from our analysis of 73 CVEs across four projects, 17.7\% of the vulnerable functions contain patch-similar code, affecting 176 test cases in $Dataset_{-irr}$. These patch-similar code segments span no more than four lines of code, primarily involving function call modifications or the addition of simple checks. Robin and PS3 correctly identify only 64\% and 44\% of these test cases, respectively, whereas \toolname{} accurately identifies 88\%, owing to its patch path verification mechanism. 
}

\begin{tcolorbox}
\textbf{Answer to RQ1.} \toolname{} can effectively identify vulnerable functions from the fixed functions with 87.8\% $TPR$, 15.2\% $FPR$ on average for the three experiments. Compared to other baselines, \toolname{} outperforms them across compilation settings with an improvement on $TPR$ and $FPR$ by 36.4\% and 54.7\%, respectively. 
\end{tcolorbox}

\begin{figure}[t]
  \centering
  \includegraphics[width=0.8\linewidth]{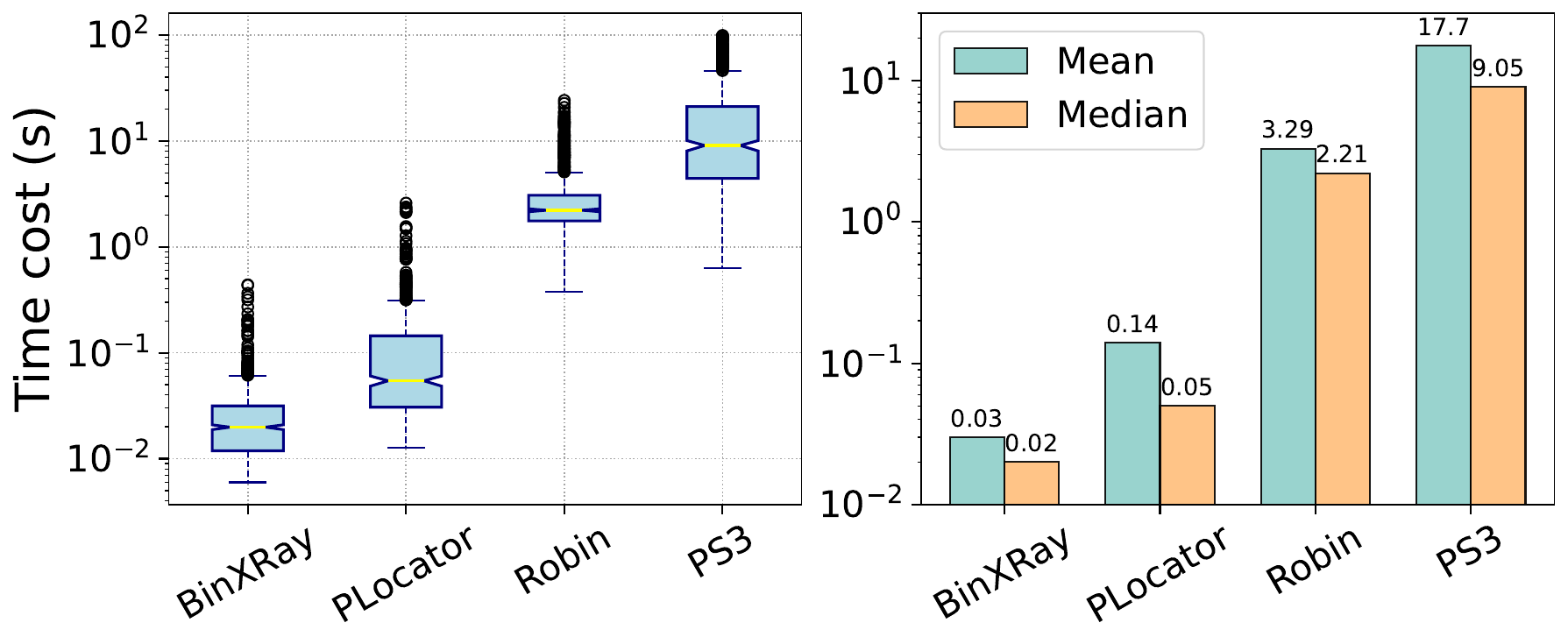}
  \caption{The time costs of different approaches. The values for Robin and PS3 are displayed on a separate scale. The left two figures present box plots for the test cases, while the right two figures illustrate the mean and median time costs.}
  \label{fig: time cost}
\end{figure}

\subsection{RQ2: Result of Vulnerability Detection Efficiency}
\cp{\Cref{fig: time cost} illustrates the time cost for a single test case across different approaches. The left figure displays box plots representing the time cost for each test case, while the right figure shows the mean and median time costs for each approach.
As observed, BinXRay exhibits exceptional speed, averaging only 0.03 seconds per test case by focusing exclusively on syntactic-level feature extraction and comparison. In contrast, Robin and PS3 require an average of 3.23 seconds and 17.7 seconds per test case, respectively, with the majority of their time spent on symbolic execution. For larger target functions exceeding 500 blocks, these methods become significantly more time-consuming due to the emulation of the target and may even fail due to unforeseen exceptions.
}

\toolname{} achieves an average detection time of 0.14 seconds per test case, maintaining consistent time efficiency regardless of function size. This remarkable efficiency is a result of the meticulously designed strategy aimed at reducing the number of candidate functions and anchor paths for matching, as outlined in~\Cref{sec: anchor path matching}. By minimizing the number of potential anchors based on their values, auxiliary information, and the distance between adjacent anchors, we effectively reduce the search space, thereby significantly lowering the number of anchor paths to match. Additionally, all features are extracted directly from the CFG, eliminating the need for extra feature generation costs and enabling a rapid matching process.

\begin{tcolorbox}
\textbf{Answer to RQ2.} \toolname{} efficiently performs vulnerability detection in a minimal amount of time with 0.14s per target function on average, providing enhanced scalability for practical application in large-scale scenarios, all while maintaining superior accuracy compared to the baselines.
\end{tcolorbox}

\subsection{RQ3: Result of Vulnerability Detection with Irrelevant Functions}
In real-world scenarios (e.g., stripped binaries), irrelevant functions cannot be disregarded and must also be identified by patch presence test approaches, which is a more challenging task that has been overlooked by prior works~\cite{Xu2020PatchBV, Zhang2018PreciseAA, Zhan2023PS3PP}. 
To assess whether the approaches perform effectively when irrelevant functions are involved, we ran the baselines and \toolname{} on $Dataset_{+irr}$ to evaluate their performance in a 1-day vulnerability detection task and calculated the metrics based on their outputs on test cases.
The results are presented in~\Cref{tab: rq3_res}. The first five rows display the results of the baselines, while the last four rows show the result for \toolname{} and the ablation study on two of its components. 

\begin{table}[]
\caption{The results of the vulnerability detection on $Dataset_{+irr}$ (\%). }
\label{tab: rq3_res}
\begin{adjustbox}{max width=\textwidth}
\begin{threeparttable}

\begin{tabular}{c|rrrrr|rrrrr|rrrrr|rr}
\hline
\hline
\textbf{Experiment} & \multicolumn{5}{c|}{\textbf{Same}} & \multicolumn{5}{c|}{\textbf{XO}}  & \multicolumn{5}{c|}{\textbf{XC}}  & \multicolumn{2}{c}{\textbf{Average}} \\ \hline
\textbf{Metric} &
  \multicolumn{1}{c}{$TPR_s$} &
  \multicolumn{1}{c}{$FPR_s$} &
  \multicolumn{1}{c}{$SR$} &
  \multicolumn{1}{c}{$TPR$} &
  \multicolumn{1}{c|}{$FPR$} &
  \multicolumn{1}{c}{$TPR_s$} &
  \multicolumn{1}{c}{$FPR_s$} &
  \multicolumn{1}{c}{$SR$} &
  \multicolumn{1}{c}{$TPR$} &
  \multicolumn{1}{c|}{$FPR$} &
  \multicolumn{1}{c}{$TPR_s$} &
  \multicolumn{1}{c}{$FPR_s$} &
  \multicolumn{1}{c}{$SR$} &
  \multicolumn{1}{c}{$TPR$} &
  \multicolumn{1}{c|}{$FPR$} &
  \multicolumn{1}{c}{$TPR$} &
  \multicolumn{1}{c}{$FPR$} \\ \hline
BinXRay             & 74.6  & 0.5  & 17.8  & 44.0 & 0.0  & 28.4 & 0.6  & 14.1  & 6.7  & 0.0  & 40.0 & 15.0 & 3.7   & 3.5  & 0.5  & 18.1              & 0.2\tnote{1}              \\
Robin               & 95.5  & 47.5 & 88.6  & 85.0 & 42.1 & 82.0 & 49.0 & 88.6  & 72.6 & 43.5 & 55.9 & 37.3 & 87.0  & 49.1 & 32.5 & 68.9              & 39.3             \\
PS3                 & 89.3  & 64.5 & 69.9  & 67.0 & 45.0 & 69.7 & 66.0 & 56.8  & 50.5 & 37.2 & 59.3 & 53.8 & 62.7  & 46.6 & 33.5 & 54.7              & 38.6             \\ \hline
PLocator$_{-Both}$  & 83.0  & 49.4 & 100.0 & 83.0 & 49.4 & 78.3 & 50.2 & 100.0 & 78.3 & 50.2 & 81.9 & 40.9 & 100.0 & 81.9 & 40.9 & 78.3              & 46.8             \\
PLocator$_{-IFF}$   & 93.0  & 22.6 & 100.0 & 93.0 & 22.6 & 87.4 & 24.2 & 100.0 & 87.4 & 24.2 & 87.1 & 15.2 & 100.0 & 87.1 & 15.2 & 89.1              & 20.7             \\
PLocator$_{-PPV}$   & 83.0  & 18.0 & 100.0 & 83.0 & 18.0 & 77.3 & 17.5 & 100.0 & 77.3 & 17.5 & 81.9 & 9.4  & 100.0 & 81.9 & 9.4  & 80.7              & 15.0             \\ \hline
\rowcolor[HTML]{C0C0C0} 
\textbf{PLocator} &
  \textbf{93.0} &
  \textbf{12.1} &
  \textbf{100.0} &
  \cellcolor[HTML]{C0C0C0}\textbf{93.0} &
  \cellcolor[HTML]{C0C0C0}\textbf{12.1} &
  \textbf{86.3} &
  \textbf{11.3} &
  \textbf{100.0} &
  \cellcolor[HTML]{C0C0C0}\textbf{86.3} &
  \cellcolor[HTML]{C0C0C0}\textbf{11.3} &
  \textbf{86.2} &
  \textbf{5.6} &
  \textbf{100.0} &
  \cellcolor[HTML]{C0C0C0}\textbf{86.2} &
  \cellcolor[HTML]{C0C0C0}\textbf{5.6} &
  \textbf{88.5} &
  \textbf{9.6} \\ \hline
  \hline
\end{tabular}%

\begin{tablenotes}
      \item[1] The low FPR of BinXRay is due to the fact that it returns ``unknown'' for most test cases, which are classified as irrelevant by default.
\end{tablenotes}
\end{threeparttable}
\end{adjustbox}

\end{table}

\paragraph{Irrelevant functions interference analysis}
\cp{As shown in~\Cref{tab: rq3_res}, the support rate of BinXRay quickly drops from 58.4\% to 17.8\% in the \emph{Same} experiment due to the large difference between the irrelevant functions and two reference functions. 
Robin and PS3 fail to identify a significant number of irrelevant functions, misclassifying them as vulnerable with average FPRs of 39.3\% and 38.6\%, respectively. This means that, for each vulnerability, they would incorrectly classify 20 and 19 test cases as vulnerable from the top 50 candidates. In contrast, \toolname{} recalls 88.5\% of the vulnerable functions with a much lower average FPR of 9.0\% across the three experiments, demonstrating its superior ability to distinguish irrelevant functions from vulnerable and fixed ones.
}

\begin{figure}[ht]
  \centering
  \includegraphics[width=\textwidth]{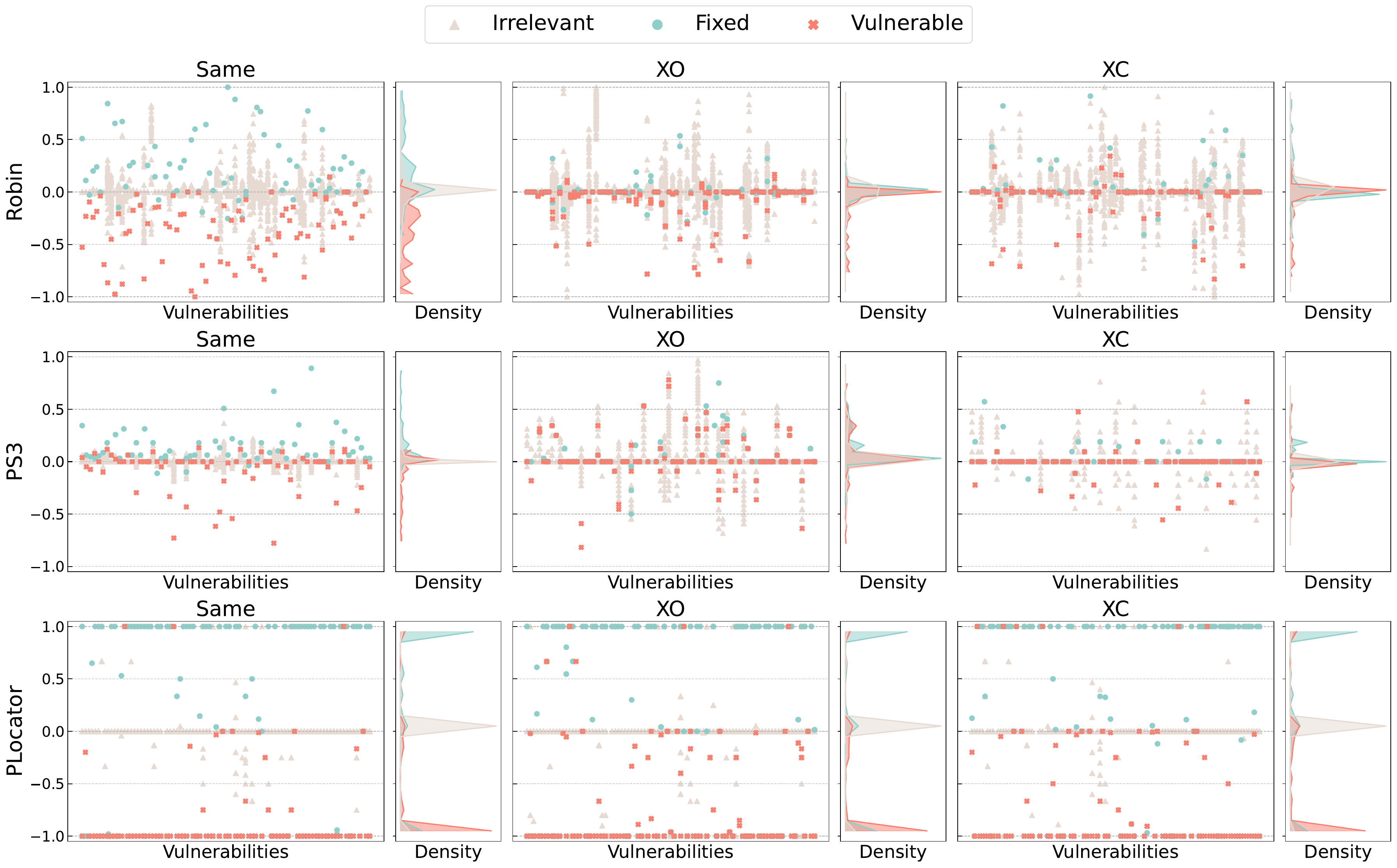}
  \caption{Test case score scatter plot and curve plot of Robin PS3, and PLocator on $Dataset_{+irr}$. The legend indicates the true category of the test case. The \emph{y-axis} refers to the score, and the \emph{x-axis} in scatter and curve refers to the different vulnerabilities and probability density of test cases, respectively.}
  \label{fig: score dist rq3}
\end{figure}

\cp{
Similar to \textbf{RQ1}, we present the score distribution of Robin, PS3, and \toolname{} as depicted in~\Cref{fig: score dist rq3} for test cases with irrelevant functions. 
\emph{Under ideal circumstances, all irrelevant test cases should yield a score of 0, indicating that they are neither vulnerable nor fixed.}
As observed, many irrelevant functions receive scores with high absolute values in both Robin and PS3 across the three experiments, even surpassing those of vulnerable and fixed functions. This makes it infeasible to distinguish vulnerable and fixed functions from irrelevant ones by setting a score threshold, as it may inadvertently filter out true positives. 
These irrelevant functions contain features (either syntactic or semantic) extracted from the reference functions, which leads to their misidentification as vulnerable or fixed. This highlights that the identification of irrelevant functions is crucial and cannot be disregarded in the 1-day vulnerability detection task. 
In contrast, \toolname{} assigns accurate scores to the majority of test cases, effectively separating fixed, vulnerable, and irrelevant test cases into three distinct groups: top (fixed, positive scores), middle (irrelevant, score $0$), and bottom (vulnerable, negative scores), as demonstrated by the curve plots in~\Cref{fig: score dist rq3} across the three experiments.
}

\paragraph{Ablation study}
As shown in the last four rows of~\Cref{tab: rq3_res}, PPV (\toolname{}$_{-IFF}$) improves TPR by 10.0\% and FPR by 55.9\%. This component distinguishes the real patch code from patch-similar code by verifying the control logic between the patch and its context, enabling the successful recall of positives previously misidentified. Additionally, many irrelevant functions that fail the patch path verification are correctly excluded, reducing false positives.
IFF (\toolname{}$_{-PPV}$) primarily enhances FPR by 68.1\% with only a slight decrease in TPR (0.4\%). This component compares the entire function rather than focusing solely on features related to the patch code and its context, proving effective, particularly when the patch path and its context are short and provide limited information.
When both components are employed, \toolname{} achieves optimal results, attaining a TPR of 88.5\% and a FPR of 9.6\%, as demonstrated in the last row of~\Cref{tab: rq3_res}.

\begin{tcolorbox}
\textbf{Answer to RQ3.} \toolname{} effectively identifies vulnerable functions within a function pool containing numerous irrelevant functions, outperforming the state-of-the-art approaches by 28.4\% in TPR and 75.0\% in FPR, highlighting its practicality for the 1-day vulnerability detection task. By integrating the two components, \toolname{} achieves optimal performance, with a 9.2\% improvement in TPR and a 79.4\% enhancement in FPR.
\end{tcolorbox}

\section{Discussion}
\subsection{Threats to Validity}
\paragraph{Internal validity}
\toolname{} relies on IDA to build the CFGs from the binaries. However, IDA may occasionally fail to generate CFG correctly or to identify the entry address of the functions in stripped binaries. To mitigate this, we can introduce other binary analysis tools (e.g., Ghidra~\cite{Ghidra}, angr~\cite{angr}) to revise the CFG and utilize some function identification technique to identify the function~\cite{Yin2018FunctionRI, Bao2014BYTEWEIGHTLT, Ye2023DARMDA}. \toolname{} depends on the BCSD tool jTrans to match the invoked functions between the reference and target functions when function names are unavailable. We found part of the functions are not recalled by jTrans due to its limitation.  It can be substituted with a more effective tool if one becomes available.

\paragraph{External validity}
We only consider four popular projects that have been tested in previous studies~\cite{Xu2020PatchBV, Yang2023TowardsPB}, which may result in some specific vulnerabilities being overlooked.
The vulnerability information gathered from NVD~\cite{NVD} may be inaccurate or outdated~\cite{Xu2021TrackingPF, Bao2022VSZZAI}. Some approaches have been proposed to mitigate it by identifying the truly vulnerable functions and locating the patch commits~\cite{Xu2021TrackingPF, Tan2021LocatingTS, Bao2022VSZZAI}. 
\toolname{} relies on the source patch file to identify the patch code block in binaries, which may not always be available, such as in cases of vulnerabilities without disclosed patch or closed-source software. As an alternative, the changed code blocks can be located by diffing two reference binary functions using existing tools (e.g., BinDiff~\cite{bindiff}) to replace the \emph{patch code mapping} step.
Currently, \toolname{} only supports the x86 architecture. When migrating to other architecture, the core idea remains unchanged while the method for extracting the two types of anchors and their auxiliary information will require adjustments with some engineering effort. For example, different architectures will utilize different calling conventions, such as x86 moving the parameters into the stack while \emph{ARM} stores the first four parameters in registers.

\subsection{Feature Selection and Usage}
\cp{The two types of key instructions selected by \toolname{} have also been utilized in previous works~\cite{Jiang2020PDiffSP, Zhan2023PS3PP, Yang2023TowardsPB}, albeit in different formats. However, \toolname{} leverages them in three different aspects:  
\begin{enumerate}[leftmargin=.25in]
    \item \toolname{} extracts constants from key instructions instead of relying on specific instruction operands and opcodes, which may vary across different compilation settings, ensuring greater robustness. For stripped binaries, \toolname{} utilize BCSD tool to recall the function calls and narrow down the search space to reduce the false positives.
    \item Beyond key instructions, \toolname{} extracts constants from sliced instructions associated with the key instructions via data dependencies, enabling the collection of more comprehensive information while mitigating noise effects to enhance detection accuracy.
    \item Unlike previous methods that overlook control logic by merely verifying the presence of features in the target function, leading to potential misidentification of patch-similar code as patched,\toolname{} separates the patch and context into individual paths and verifies their logical relationships, effectively distinguishing real patch code from patch-similar code.  
\end{enumerate}
}

\begin{figure}[ht]
  \centering
  \includegraphics[width=0.8\linewidth]{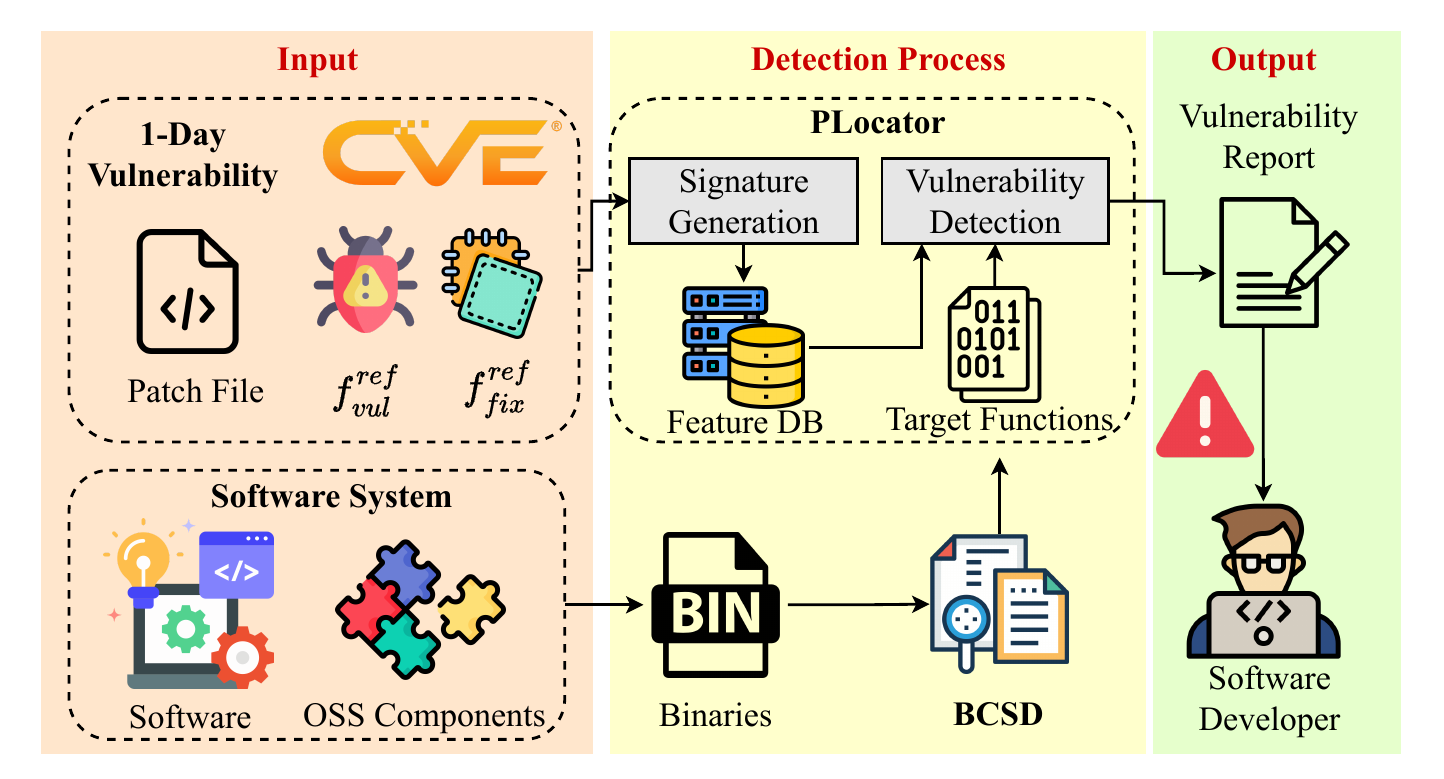}
  \caption{The application scenario of \toolname{}.}
  \label{fig: application scenario}
\end{figure}

\subsection{Application Scenario of \toolname{}}
\Cref{fig: application scenario} shows the application scenario of \toolname{}. For a new disclosed or existing vulnerability with the patch file and two reference functions ($f_{vul}^{ref}$ and $f_{fix}^{ref}$), software developers can utilize \toolname{} to verify whether the vulnerability exists in their software system. 

Initially, \toolname{} takes the 1-day vulnerability information as input and generates signatures, which are stored in a feature database to avoid redundant generation. Second, leveraging existing BCSD tools such as jTrans, \toolname{} processes the features of the vulnerability and tests the target functions provided by the BCSD to generate a vulnerability report. Ultimately, based on the vulnerability report, software developers can verify the identified vulnerable functions and determine the presence of the vulnerability within the software system.

\section{Related Work}
This section reviews related works closely aligned with 1-day vulnerability detection, encompassing two categories: binary code similarity detection and patch presence testing.

\subsection{Binary Code Similarity Detection}
With a known source function, binary code similarity detection (BCSD) compares two or more pieces of binary code to identify the similarities between the source and the target. Vulnerability detection is one of its application scenarios, it also supports malware detection~\cite{Krgel2005PolymorphicWD, bruschi2007code, cesare2013control} and software plagiarism detection~\cite{Luo2014SemanticsbasedOB, Luo2017SemanticsBasedOB}. 
In earlier years, most BCSD approaches primarily relied on syntax and code structures to compare two functions, struggling to mitigate interference from compilation settings. 
For example, DCC~\cite{Sbjrnsen2009DetectingCC} normalizes instructions and segments the function into pieces using a sliding window to calculate similarities.
~\cite{David2014TraceletbasedCS} decomposes the CFG into consecutive chunks of code segments called tracelets and computes function similarity based on these tracelets.
DiscovRE~\cite{eschweiler2016discovre} filters candidate functions using statistical numeric features in blocks and computes similarity based on the structure of CFGs.
With the development of AI and NLP techniques, many approaches proposed to combine this new technique in BCSD, which surpasses traditional methods in terms of scalability and accuracy. 
Gemini~\cite{xu2017neural} proposes the Attributed Control Flow Graph (ACFG), which attaches a set of attributes to each CFG block, and uses a graph embedding to enable training of graph neural networks.
VulSeeker~\cite{gao2018vulseeker} enhances the ACFG of Gemini by incorporating data flow, i.e., the read and write memory operations.
jTrans~\cite{Wang2022jTransJT} is the state-of-the-art BCSD tool, which uses the jump instructions to split binary code and the powerful BERT model~\cite{Devlin2019BERTPO} to extract function semantics.

BCSD is often a prerequisite for 1-day vulnerability detection due to its high efficiency and recall. 
However, it fails to capture subtleties in patches and achieves low precision when the number of candidate functions is large. 
Many irrelevant functions similar to the vulnerable one are involved, causing many false positives if used alone.

\subsection{Patch Presence Test}
Patch Presence Test identifies vulnerabilities in binary by verifying the presence/absence of the patch. Syntactic-based methods focus on the syntactic-level modifications introduced by the patch and extract corresponding features for matching. 
Fiber~\cite{Zhang2018PreciseAA} associates the changes introduced by the patch with binary instructions and matches them in the target binary with the CFG topology and abstract syntax trees.
BinXRay~\cite{Xu2020PatchBV} maps the changed patch blocks in binary and extracts the traces at the block level to compare the target and two reference functions. 
PatchDiscovery~\cite{Xu2023PatchDiscoveryPP} prioritizes the more representative blocks, namely, those with more changed instructions, to mitigate the over-fitting problem encountered in BinXRay.
Semantic-based methods such as PDiff~\cite{Jiang2020PDiffSP}, PS3~\cite{Zhan2023PS3PP}, and Robin~\cite{Yang2023TowardsPB} all utilize symbolic execution to simulate the program and extract different types of semantic features as the features during the execution. After that, they verify the presence of the signature (PS3) or calculate the similarities between the target and two reference functions (PDiff and Robin) to determine whether the binary has been patched or not.

\cp{There are also some patch presence test works on Java. BScout~\cite{Dai2020BScoutDW} is a tool for detecting patches in Java binaries without using signatures. BScout introduces techniques to align source code with bytecode and verify patch semantics across the entire executable. Ppt4J~\cite{Pan2023Ppt4jPP} detects patches by analyzing semantic features from Java source code and bytecode, comparing patched and unpatched binaries through feature matching. It identifies relevant changes, evaluates patch presence using a voting system, and ensures precise detection of subtle patch modifications. 
}

These patch presence test works focus on the code changes introduced by patches and thus are superior to BCSD in distinguishing vulnerable and fixed functions.
Yet, they fail to detect vulnerabilities effectively in practical environments due to interference mentioned in~\Cref{sec: introduction}. 
In contrast, our approach successfully mitigates the interference by concentrating on the stable values across compiler options and utilizing the patch contexts to accurately identify vulnerabilities.

\section{Conclusion}
1-day vulnerability detection is a critical and challenging task. Existing methods face difficulties in addressing interference from diverse compilers, optimizations, irrelevant functions, and patch-similar code blocks. We introduce \toolname{}, a novel approach leveraging patches and their context, represented through stable values, to accurately locate patch code and detect vulnerabilities in binaries with exceptional speed. Comprehensive experiments on two datasets (with and without irrelevant functions) demonstrate that \toolname{} significantly outperforms state-of-the-art approaches by a considerable margin.

\bibliographystyle{ACM-Reference-Format}
\bibliography{reference}

\end{document}